\documentclass[journal]{IEEEtran}
\usepackage{amssymb, amsmath}
\usepackage{bm}
\usepackage{color}
\usepackage{graphicx}
\usepackage{cite}
\usepackage{booktabs}
\usepackage{multirow}
\usepackage{threeparttable}
\usepackage{enumerate}
\usepackage[english]{babel}
\usepackage[ruled, lined, linesnumbered, commentsnumbered, longend]{algorithm2e}
\usepackage{multicol}
\usepackage{commath}
\usepackage{makecell}
\usepackage{blindtext}
\usepackage{soul}
\usepackage{comment}
\usepackage{tabularx}
\usepackage{xcolor}

\newcommand{\vect}[1]{\mathbf{#1}}
\newcommand{\matr}[1]{\mathbf{#1}}

\newcommand{\tran}{\mathsf{T}}
\newcommand{\est}[1]{\hat{#1}}
\newcommand{\pluseq}{\mathrel{+}=}

\newtheorem{theorem}{Theorem}

\newtheorem{remark}{Remark}

\makeatletter
\renewcommand*\env@matrix[1][\arraystretch]{%
  \edef\arraystretch{#1}%
  \hskip -\arraycolsep
  \let\@ifnextchar\new@ifnextchar
  \array{*\c@MaxMatrixCols c}}
\makeatother

\graphicspath{{Figures/}}

\begin{document}
\title{Modularized Bilinear Koopman  Operator for Modeling and Predicting Transients  of   Microgrids}
\author{Xinyuan Jiang, 
        Yan~Li,~\IEEEmembership{Senior Member,~IEEE,} and Daning Huang
\thanks{X. Jiang and Y. Li are with the Department of Electrical  Engineering, The Pennsylvania State University, University Park, PA 16802, USA (e-mail: yql5925@psu.edu).}
\thanks{ D. Huang is with the Department of Aerospace  Engineering, The Pennsylvania State University, University Park, PA 16802, USA (e-mail: daning@psu.edu).
}
}

\maketitle

\begin{abstract}

Modularized Koopman Bilinear Form (M-KBF) is presented to model and predict the transient dynamics of microgrids in the presence of  disturbances.
As a scalable data-driven approach, M-KBF divides the identification and prediction of the high-dimensional nonlinear system into the individual study of subsystems; and thus, alleviating the difficulty of intensively handling high volume data and overcoming the curse of dimensionality.
For each subsystem, Koopman bilinear form is applied to efficiently identify its model by developing eigenfunctions via the extended dynamic mode decomposition method with an eigenvalue-based order truncation.
Extensive tests show that M-KBF can provide accurate transient dynamics prediction for the nonlinear microgrids and verify the plug-and-play modeling and prediction function, which offers a potent tool for identifying high-dimensional systems.
The modularity feature of M-KBF enables the provision of fast and precise prediction for the microgrid operation and control, paving the way towards online applications.

\end{abstract}

\begin{keywords}
Modularized Koopman bilinear form (M-KBF), 
data-driven modeling,
Koopman eigenanalysis,
extended dynamic mode decomposition (EDMD),
transient dynamics prediction,
microgrids,
distributed energy resources (DERs).
\end{keywords}
\section{Introduction}

\PARstart{D}{istributed} 
Energy Resources (DERs), such as photovoltaic (PV) and wind power, is seen as a great opportunity to achieve the target of modernizing the power systems. 
Microgrids have been developed to integrate those DERs. 
Considering most  DERs are integrated into microgrids through power-electronic interfaces,  the system’s inertia is significantly reduced. 
Consequently, microgrids are  sensitive 
to disturbances like PV fluctuations and load disturbances; and thus, it is of primary importance to  investigate the transient dynamics of  microgrids  subjected to  disturbances.

Efforts on the modeling and control of transient dynamics since the 1980s mainly focused on the mathematical model-based simulation and analysis~\cite{sobbouhi2021transient}.
Despite that several control approaches have been developed for  stabilizing the system during transients, it is still elusive  that \emph{how to analyze and predict the system's transient dynamics  when an accurate system model is unavailable.} The rapidly developing machine learning and artificial intelligence technologies~\cite{li2021determination,
adibi2019reinforcement,
madani2021data} provide a potent means to resolve this challenge from the data-driven perspective.


Recognizing the fact that  microgrid is a quintessentially nonlinear dynamical system, there are several existing data-driven approaches to identify the transient dynamics model of a nonlinear system through its operating data. They can fall into three major categories: (1) linear models as local linearization of a globally nonlinear system, (2) nonlinear models to directly capture the global nonlinearity, and (3) linear models in the embedded space to reproduce the global nonlinearity.


First, 
the local linear models are the most commonly used system identification method and develop a high-order linear system with input and output to approximate the system dynamics near an equilibrium point.
From the perspective of linear system identification~\cite{a299a76f179d4ebba7c9872e1aced639}, existing methods include Prony analysis~\cite{119247, 5275444} and state space methods, e.g., 
Minimal Realization algorithm~\cite{260898}, 
Eigenvalue Realization Algorithm (ERA)~\cite{780912}, 
Matrix Pencil method~\cite{6024892}, Hankel Total Least Squares (HTLS)~\cite{4410548},
subspace identification~\cite{9609618},
Dynamic Mode Decomposition (DMD)~\cite{6982236}, etc. 
Beyond the linear system perspective, 
there are nonparametric spectrum estimation such as Welch periodgram~\cite{stoica2005spectral} and parametric methods, including Yule-Walker~\cite{anderson2005bootstrap}, 
Frequency Domain Decomposition~\cite{liu2008oscillation}, etc.
These aforementioned approaches generate accurate model for linear systems given sufficient system responses and are straightforward to perform with guaranteed convergence. 
However, the linear system methods do not have the extrapolation ability due to their nature of local linearization; and thus, it is not suitable to directly apply them to identify microgrid systems that are typically nonlinear for the entire operating envelope.

Second, several data-driven methods have also been developed to identify a nonlinear system to capture the transient dynamics globally over the entire state space.  Representative methods include Taylor series~\cite{8598677}, 
Volterra series~\cite{1546078},
Sparse Identification (SINDy)~\cite{brunton2016discovering}, 
etc. 
These methods can theoretically identify an accurate model if the correct nonlinear terms are used.  However, their implementation usually result in a non-convex optimization formulation and need a large computational effort; and the identified models are limited by the high algorithmic complexity in the prediction stage, which poses a challenge for control applications.

Third, to reconcile the dilemma between local linearization and global nonlinearity, the global linearization method, represented by the Koopman operator theory, has been seen as a promising paradigm for the data-driven modeling and control of nonlinear systems~\cite{mezic2005spectral,mezic2004comparison,mezic2021koopman,BEVANDA2021197,budivsic2020koopman} and has been widely used in the domain of power systems~\cite{susuki2011nonlinear,korda2018power,sharma2021data}.
The Koopman theory states that a nonlinear dynamical system can be represented using an infinite-dimensional linear operator, that is characterized by an infinite set of eigenvalues and eigenfunctions, or eigenpairs, on the so-called embedded space.
Using a converged, truncated set of eigenpairs, the nonlinear dynamics can be represented by a finite-dimensional linear system on a Koopman invariant embedded subspace with the eigenfunctions as new coordinates.
Compared to the traditional nonlinear system identification models, the Koopman-based model evolves linearly in the embedded space, leading to a simplified representation of the original nonlinear system that is amenable for efficient control applications.

In the Koopman theoretic framework, the data-driven modeling centers on the identification of a set of eigenpairs that is sufficiently comprehensive to represent the nonlinear dynamics linearly in the embedded subspace and reconstruct it in the original state space.  The eigenfunctions are usually constructed as functionals of the observable functions of the state variables in the nonlinear dynamics.
The existing methods for identifying Koopman eigenfunctions can generally be divided into two types.
The first type is the Extended Dynamic Mode Decomposition (EDMD) and its variants, where a predetermined set of observable functions is given~\cite{williams2015data}.
The observable functions lift the original state in the training data into an embedded space, on which a linear system identification method such as DMD  can be applied to identify a linear model as well as a Koopman invariant subspace spanned by the correspondingly approximated Koopman eigenfunctions.
The set of observable functions are required to approximate a Koopman invariant subspace as much as possible; otherwise, the identified model may be no better than a local linear model. 
Moreover, the EDMD-type methods are usually limited by the curse of dimensionality, as the number of observable functions may grow exponentially with the number of states, which makes this type of method infeasible to directly use for identifying high-dimensional nonlinear dynamical systems like power systems and microgrids.
The second type of Koopman methods directly solves for Koopman eigenfunction basis to build the linear system~\cite{folkestad2020extended,korda2020optimal}.
This type of methods exploit the connection of the Koopman operators to state-space geometry either within a basin of attraction or off-attractor. And hence, they are limited in the case where the state-space geometry is altered by inputs.




To predict transient dynamics in the Koopman framework, inputs that represent the disturbances and cause the transient response need to be included in the linear evolution of eigenfunctions.
Previously, the nonlinear system's inputs  were added to the Koopman model linearly, which resulted in a linear system with inputs~\cite{korda2018power}.
However, this assumption applied to the function space is not consistent with how the inputs influence the dynamics of the original nonlinear system in the state space, which significantly limits the applicable range of Koopman model.
Recently, Koopman model called Koopman Bilinear Form (KBF) in the context of control-affine systems were proposed in~\cite{goswami2021bilinearization,peitz2020data}, where the bilinear input has a precise connection to the original state-space input.


In this paper,  a Modularized Koopman Bilinear Form 
(M-KBF) is developed as a scalable data-driven modeling   approach for efficiently predicting the transient dynamics of microgrids integrated with power-electronics-interfaced DERs. 
The data-driven modeling of a whole  microgrid system is divided into the separate identification of backbone subsystem and DERs dynamics. 
KBF is used  to develop a precise data-driven model for each DER, since the typical power-electronics-interfaced DERs have been determined to be control-affine to the input disturbance from the current or voltage at the DER integration point. 
The identified  KBF modules for DERs are then integrated into the backbone subsystem to assemble the entire data-driven model for the whole microgrid.
The novelties of the presented work are summarized below.
\begin{enumerate}
    \item M-KBF unlocks the potential of KBF for high-dimensional nonlinear systems by overcoming the curse of dimensionality. It develops the data-driven model of high-dimensional systems by first breaking it into several manageable subsystems that each is modeled by a  KBF model and then connecting the subsystems via the accurately modeled inputs. 
    \item M-KBF offers modularity in both data-driven modeling and prediction. It does not need to send high volume data to a central computational resource; and thus, issues caused by data acquisition like data privacy and poisoning   can be alleviated or fully avoided. Meanwhile, local distributed computational resources can be exploited for identifying  large-scale dynamical systems.
    \item The modularity design enables M-KBF with the plug-and-play versatility. The KBF models for subsystems can be built from a relatively small amount of training data of only the subsystem under consideration. Moreover, the identified KBF models for subsystems like DERs can be repeatedly used for other  systems,  which is advantageous over previous data-driven methods in the Koopman framework that focus on modeling the whole system. 
    \item M-KBF ensures accurate predictions because each identified  KBF models are  independent to the initial choice of the global DQ reference frame, which is found to be important for nonlinear system identification. It is realized by introducing random DQ frames to the measurement data set.  This property signifies a major difference to the linear data-driven models.
\end{enumerate}

Besides, considering the fact that the voltage and current usually respond to input change with no time delay,  a modification to the explicit solution procedure for the connected data-driven model  is also presented to reduce its intrinsic time delay effect on prediction.
The prediction obtained from M-KBF can potentially be applied to provide guidance for system control, such as  performing model predictive control, which is the authors' next work. 

The reminder of the paper is organized as follows.
Section II establishes the modularized Koopman bilinear form.
Section III introduces the determination of eigenfunctions through the EDMD method with order truncation.
Section IV presents the prediction of transient dynamics based on M-KBF.
In Section V, tests on a microgrid system verify the effectiveness and efficiency of the presented method in modeling and predicting transients.
Conclusions are drawn in Section VI.

\section{Modularized Koopman Bilinear Form}

The essential idea of M-KBF is to decouple a large system into several smaller subsystems, identify the data-driven model for each subsystem separately through KBF, and then integrate them into the network model with inputs and outputs of bus voltages or currents.
Considering the transient dynamics of microgrids is dominated by the nonlinear dynamics of DERs while the grid can be modeled as a set of linear constraints between node currents and voltages, the data-driven modeling of the whole system can be separated into identifying the grid's algebraic constraint and the dynamics of DER subsystems.
The main advantages of the presented M-KBF are that the  modeling and prediction architecture is scalable and the identified model is linear or control-affine in the lifted space so linear system identification methods can be leveraged ~\cite{sharma2021data}.


\subsection{Identification of Hybrid Network Parameter Matrix}

Taking into account the different functions of grid-forming and grid-following DERs in the microgrid system~\cite{kettner2021harmonic}, the  network that connects DERs and power loads can be represented by a hybrid network parameter matrix~\cite{kettner2017properties}, as given in (\ref{Eq_Hmatrix}), which gives the constraint between node currents and  voltages. 
\begin{equation} \label{Eq_Hmatrix}
    \begin{bmatrix}
        \matr{I}_{\mathcal{P}_1} \\
        \matr{V}_{\mathcal{P}_2}
    \end{bmatrix} = \matr{H} \begin{bmatrix}
        \matr{V}_{\mathcal{P}_1} \\
        \matr{I}_{\mathcal{P}_2}
    \end{bmatrix},
\end{equation}
where 
the partition $\mathcal{P}_1$ is the set of nodes connecting to the grid-forming DERs, and the partition $\mathcal{P}_2$ is the set of nodes connecting to the grid-following DERs and/or power loads. 
Let $\mathcal{N}$ be the set of microgrid nodes, then
$\mathcal{P}_1 \cup \mathcal{P}_2 \subseteq  \mathcal{N}$. 
The goal is to find a hybrid network parameter matrix $\matr{H}$ that satisfies (\ref{Eq_Hmatrix}).


Specifically, the grid-forming DER inverters take node current as input and node voltage as output in (\ref{Eq_Hmatrix}), while the grid-following inverters take node voltage as input and node current as output. 
The  power loads have the same input and output definitions as grid-following DERs. 
Therefore, the network has complementary input and output ports at the DER inverter nodes, leading to the hybrid network parameter matrix $\matr{H}$.



The hybrid network parameter matrix can be obtained by using two methods, namely by obtaining the matrix analytically from the admittance matrix if it is known, or by solving a least-squares problem based on  the measurement  of node voltages and currents. Note that both methods need to meet the condition given in Remark \ref{condition} to guarantee that $\matr{H}$ exists.
To  derive $\matr{H}$ from the admittance matrix, there are two  steps~\cite{kettner2017properties}, including
the elimination of nodes with zero current injections from generations or loads (e.g., Kron reduction~\cite{dorfler2012kron}) and the switching of a set of node currents and voltages (for partition $\mathcal{P}_2$). 
To solve for the hybrid network parameter matrix from measurements of node voltages and currents, it is typical to formulate (\ref{Eq_Hmatrix}) into a least-squares problem, which is well posed given the existence of $\matr{H}$ (Remark \ref{condition}).

\begin{remark}[Sufficient condition for the existence of $\matr{H}$] \label{condition}
In a connected network, assume all the branches are not electromagnetically coupled and have nonzero admittance. If all lines are passive, i.e., having strictly positive real part in line admittance, then the hybrid network parameter exists~\cite{kettner2017properties}. The existence follows from the ability to transform the admittance matrix through the two steps described above. If, in particular, no shunt elements (shunt capacitors or constant impedance loads) are included in the network, then the hybrid matrix always exists for a connected network.

\end{remark}

\subsection{Bilinear Koopman Surrogate Model for DERs}

Koopman Bilinear Form is a global bilinearization of control-affine nonlinear systems in an operator theoretic view.
Each grid-forming controlled DER with node current inputs or the grid-following DER with node voltage inputs can be expressed in the following form of control-affine system, with  Vf controlled DER  provided as an example in the Appendix. 
\begin{subequations}\label{E:control_affine_system}
\begin{align}
    \dot{\vect{x}} &= \vect{f}(\vect{x}) + \sum_{i=1}^m \vect{g}_i(\vect{x}) u_i, \\
    \vect{y} &= \matr{C}\vect{x}, \label{E:output}
\end{align}
\end{subequations}
where $\vect{x} \in \mathbb{X} \subseteq \mathbb{R}^n$ is the state vector for one DER, $\vect{u} = [u_1 \ldots u_m]^{\tran} \in \mathbb{R}^m$ is the input vector, $\vect{y} \in \mathbb{R}^2$ is the output vector, $\matr{C} \in \mathbb{R}^{2{\times}n}$ is the output matrix,
$\vect{f}: \mathbb{X} \to \mathbb{X}$ and $\vect{g}_i: \mathbb{X} \to \mathbb{X}$ are the flow and control vector fields for the two typical  inverter controls. 
The KBF for identifying DER dynamics are then introduced from the following three aspects. 

\subsubsection{Observable Function}
An observable function $\varphi: \mathbb{X} \to \mathbb{C}$ is a complex-valued function of the state vector $\vect{x}$. Let $\mathcal{F}$ be the space of all possible observable functions so that $\varphi \in \mathcal{F}$.
Let $\mathbf{\Phi}_{\vect{u}}(t,\vect{x}_0)$ be the flow map of the system (\ref{E:control_affine_system}) at time $t > 0$ starting from an initial condition $\vect{x}_0$ with input $\mathbf{u}$.
The time-varying observable  $\psi(t,\vect{x}) \triangleq \left.\varphi(\vect{x})\right|_{\vect{x} = \mathbf{\Phi}_{\vect{u}}(t,\vect{x}_0)}$ of the system (\ref{E:control_affine_system}) is the solution of the following partial differential equation,  which lays the foundation for bilinearization.
\begin{equation}\label{E:evolution_PDE}
\begin{split}
    \frac{\partial \psi}{\partial t} &= L_{\vect{f}}\psi + \sum_{i=1}^{m} u_i L_{\vect{g}_i} \psi, \\
    \psi(0,\vect{x}) &= \varphi(\vect{x}),
\end{split}
\end{equation}
where $L_{\vect{f}} \triangleq \vect{f} \cdot \nabla$, $L_{\vect{g}_i} \triangleq \vect{g}_i \cdot \nabla,\,i = 1,\ldots,m$ are the Lie derivatives~\cite{7384725} with respect to the drift and control vector fields $\vect{f}(\vect{x})$ and $\vect{g}_i(\vect{x})$, which are linear operators on $\mathcal{C}^1(\mathbb{X})$.

\subsubsection{Koopman Operator}
The Koopman operator is defined for the system (\ref{E:control_affine_system}) 
with zero inputs,
\begin{equation}\label{E:autonomous_system}
    \dot{\vect{x}} = \vect{f}(\vect{x})
\end{equation}

Assume ${\mathbf{\Phi}}(t,\vect{x}_0)$ is the flow map of (\ref{E:autonomous_system}) for the time $t > 0$, the continuous time Koopman operator is defined as $\mathcal{K}^t: \mathcal{F} \to \mathcal{F}$ such that,
\begin{equation}\label{E:Koopman_operator}
    (\mathcal{K}^t\varphi)(\cdot) = \varphi \circ \mathbf{\Phi}(t,\cdot),
\end{equation}
where $\circ$ represents function composition. Based on (\ref{E:Koopman_operator}), we can see the Koopman operator is linear, namely,
\begin{equation}\label{E:linearity}
\begin{aligned}
    \mathcal{K}^t (\alpha_1 \varphi_1 + \alpha_2 \varphi_2) &= \alpha_1\varphi_1 \circ \mathbf{\Phi}(t,\cdot)  + \alpha_2 \varphi_2 \circ \mathbf{\Phi}(t,\cdot)\\
    &=\alpha_1\mathcal{K}^t  \varphi_1 + \alpha_2 \mathcal{K}^t \varphi_2.
\end{aligned}
\end{equation}

Therefore, the Koopman operator can be characterized by its eigenvalues and eigenfunctions,
\begin{equation}\label{E:eigen_linear}
    \mathcal{K}^t \phi = e^{t\lambda} \phi,
\end{equation}
where $\lambda \in \mathbb{C}$  is the Koopman eigenvalue and  $\phi \in \mathcal{F}$ is the corresponding eigenfunction. 


Through the definition of Lie derivatives, the infinitesimal generator of the Koopman operator is equal to the Lie derivative of the drift vector field, i.e., $L_{\vect{f}} = \lim_{t \to 0} \frac{\mathcal{K}^t - I}{t}$, where $I$ is the identity operator, so $L_{\vect{f}}$ is also referred to as the Koopman generator. 
The eigenvalue and eigenfunction relation in (\ref{E:eigen_linear}) can also be expressed in terms of the Lie derivative,
\begin{equation}\label{E:eigenfunction}
    L_{\vect{f}} \phi = \lambda \phi.
\end{equation}

From (\ref{E:eigenfunction}) and the definition of $L_{\vect{f}}$, for any two eigenpairs $(\lambda_1,\phi_1)$ and $(\lambda_2,\phi_2)$, we have
\begin{equation}\label{E:eigenvalue_group}
    L_{\vect{f}}(\phi_1 \cdot \phi_2) = (\lambda_1 + \lambda_2) (\phi_1 \cdot \phi_2).
\end{equation}
Thus, there exist  infinitely many eigenfuncions and eigenvalues for the Koopman operator.

\subsubsection{Bilinearzation}
The goal of bilinearzation of (\ref{E:control_affine_system}) is to choose a set of observable functions,
\begin{equation}
    T(\vect{x}) = [\varphi_1(\vect{x}),\varphi_2(\vect{x}),\, \ldots,\, \varphi_N(\vect{x})]^{\tran},
\end{equation}
such that their evolution over time is that of a bilinear system (\ref{E:KBF}), which mirrors (\ref{E:evolution_PDE}) with the observable function basis $T(\vect{x})$.
\begin{equation}\label{E:KBF}
    \dot{\vect{z}} = \matr{A}\vect{z} + \sum_{i=1}^m u_i \matr{B}_i \vect{z},\quad \vect{z}(0) = T(\vect{x}_0),
\end{equation}
where $\matr{A}$ and $\matr{B}_i$ will be determined in Section III.
The condition for bilinearizability into (\ref{E:KBF}) with the state embeddings $T(\vect{x})$ being the eigenfunctions of $L_{\vect{f}}$ is provided in Theorem 1 from~\cite{goswami2021bilinearization}. 
And furthermore, bilinearizability condition with finite eigenfunction embeddings~\cite{goswami2021bilinearization} is presented in the following Theorem for completeness.

\begin{theorem}
\label{Thm_finite_bilinearizable}
    If a set of Koopman eigenfunctions $\{\phi_1,\phi_2,\ldots,\phi_n\},\, n \in \mathbb{N}$ of the unactuated system forms an invariant subspace of $L_{\vect{g}_i},\, i = 1,\ldots,m$, then the system (\ref{E:control_affine_system}) is bilinearizable with an $n$ dimensional state space.
\end{theorem}


Given that the condition of Theorem~\ref{Thm_finite_bilinearizable} is satisfied, the control-affine DER subsystem (\ref{E:control_affine_system}) is bilinearizable with a finite number of eigenfunction embeddings. 
Next, we will obtain a finite-dimensional bilinear model (\ref{E:KBF}) by modifying the Extended Dynamic Mode Decomposition (EDMD) algorithm to include blinear inputs.
\section{Data-Driven Identification of KBF}

The data-driven identification of the KBF equivalent model (\ref{E:KBF}) of DER subsystems is carried out through a least-squares formulation using a predetermined set of observable functions. 
The eigenfunction embeddings and the KBF system are identified simultaneously from the state and input measurements of the DER subsystems.

The general rule of selecting a dictionary of observable functions is that these functions should span a rich subspace so that a certain set of eigenfunctions can be approximated by their projections onto this subspace. 
There are several choices of observable functions~\cite{bruce2019koopman}, including monomials, radial basis functions, Hermite polynomials, and Chebyshev polynomials.
And in this work we choose monomial functions as the observable dictionary.
In this paper, we use EDMD with additional bilinear inputs to identify the KBF model (\ref{E:KBF}). 
As will be shown in the subsection A, the KBF model identified in this way inherently satisfies the bilinearizability condition in Theorem~\ref{Thm_finite_bilinearizable}.


\subsection{Approximated Eigenfunctions}

Given the conditions in Theorem~\ref{Thm_finite_bilinearizable}, the original system (\ref{E:control_affine_system}) can be identified by the surrogate KBF model (\ref{E:KBF}) in the coordinates of a finite set of eigenfunctions of $L_{\vect{f}}$. One option to approximate such eigenfunctions of $L_{\vect{f}}$ is by applying EDMD on the sampled trajectories of the system with zero inputs. 
Since the inputs to DER subsystems include node current or voltage that cannot be held constant during system transients, it is not feasible to independently identify the eigenfunctions using EDMD.
Instead, the invariant eigenfunction embeddings and the bilinear system can be identified together using a predetermined dictionary of observable functions. 

The requirement on the observable functions is that their span approximates an invariant eigenspace of $L_{\vect{f}}$ w.r.t (\ref{E:evolution_PDE}) that is defined in Theorem~\ref{Thm_finite_bilinearizable}. Then given the KBF system (\ref{E:KBF}) and the observable functions $T(\vect{x})$, for each left eigenvector $\vect{w}_i$ and eigenvalue $\lambda_i$ of $\matr{A}$, $\hat{\phi}(\vect{x}) = \vect{w}_i^* T(\vect{x})$ is proved to be an eigenfunction of $L_{\vect{f}}$ with eigenvalue $\lambda_i$~\cite{proctor2016dynamic}. Assume that all eigenvalues of $\matr{A}$ are distinct. Denote the left eigenvector matrix $\matr{W}$ and eigenvalue matrix $\matr{D}$ of $\matr{A}$ such that $\matr{W}^* \matr{A} = \matr{D}\matr{W}^*$. Then the approximated eigenfunction embeddings $\tilde{\vect{z}} = \matr{W}^* \vect{z} = \matr{W}^*T(\vect{x})$ is a linear transformation of the predetermined observable function embeddings, which gives us the following KBF system with the eigenfunction embedded state,
\begin{equation}\label{E:KBF_eigenfunction_coordinates}
    \dot{\tilde{\vect{z}}} = \matr{D} \tilde{\vect{z}} + \sum_{i=1}^m u_i \tilde{\matr{B}}_i \tilde{\vect{z}},\quad \tilde{\vect{z}}(0) = \matr{W}^*T(\vect{x}_0),
\end{equation}
where $\tilde{\matr{B}}_i = \matr{W}^* \matr{B}_i (\matr{W}^*)^{-1}$. 



\subsection{Discretization of KBF for Identification}

Since the measurement data are sampled data, the continuous-time KBF system (\ref{E:KBF}) needs to be discretized to provide the discrete-time template for identification. Again, we assume that the span of the observable functions $T(\vect{x})$ is a subspace of invariant eigenspace of $L_{\vect{f}}$. 
Assume the sampling period is $\Delta t$ and that the inputs $u_i(t)$ are constant within each time step (zero-order hold). The KBF system is linear time-invariant  with the solution (\ref{E:LTI_solution}).
\begin{equation}\label{E:LTI_solution}
    \vect{z}(t+\Delta t) = \exp\Bigg[\Bigg(\matr{A} + \sum_{i=1}^m u_i(t+\Delta t)\matr{B}_i\Bigg) \Delta t \Bigg]\vect{z}(t)
\end{equation}

Then, by expanding (\ref{E:LTI_solution}) into Taylor series and taking the first order approximation on $\Delta t$, we can get the discretized KBF model in explicit (\ref{E:discrete_explicit}) and implicit (\ref{E:discrete_implicit}) forms:
\begin{subequations}\label{E:discrete}
\begin{align}
    \vect{z}_{k+1} &= \matr{A}^d\, \vect{z}_k + \sum_{i=1}^m u_{i,k+1} \matr{B}_i^d\, \vect{z}_k + \mathcal{O}(\Delta t^2), \label{E:discrete_explicit} \\
    \vect{z}_{k+1} &= \matr{A}^d\, \vect{z}_k + \sum_{i=1}^m u_{i,k+1} \matr{B}_i^d\, \vect{z}_{k+1} + \mathcal{O}(\Delta t^2), \label{E:discrete_implicit}
\end{align}
\end{subequations}
where $\matr{A}^d = \exp(\matr{A} \Delta t),\, \matr{B}_i^d  = \matr{B}_i \Delta t$.

\subsection{Least-Squares Formulation}

Suppose that a suitable dictionary of observable functions $T(\vect{x}) \in \mathbb{R}^q$ is selected, we can identify the discretized KBF models (\ref{E:discrete}) using least-squares formulation similar to EDMD~\cite{proctor2016dynamic}, which gives us the approximated eigenfunction coordinates and the KBF (\ref{E:KBF}). In the following, we assume that the explicit model (\ref{E:discrete_explicit}) is used, while the implicit model (\ref{E:discrete_implicit}) can be obtained in a similar way.

Assume that the measurement data is collected from a single sampled trajectory which includes many transient responses of the DER under investigation. Denote the state and input vector data as $\vect{x}_k,\, \vect{u}_k$ for $k = 1,\, \ldots\,,\, N$. We will organize the measurement data into the following matrices:
\begin{subequations}
\begin{align}
    \matr{X}_1 &= \Big[T(\vect{x}_1),\, T(\vect{x}_2),\, \ldots \,,\, T(\vect{x}_{N-1})\Big], \\
    \matr{X}_2 &= \Big[T(\vect{x}_2),\, T(\vect{x}_3),\, \ldots \,,\, T(\vect{x}_{N})\Big], \\
    \matr{\Gamma}_i &= \Big[u_{i,2} T(\vect{x}_1),\, u_{i,3} T(\vect{x}_2),\, \ldots\,,\, u_{i,N} T(\vect{x}_{N-1})\Big].
\end{align}
\end{subequations}

Then, a least-squares problem based on the discrete-time model (\ref{E:discrete_explicit}) can be formulated  in (\ref{E:ls_KBF}).
\begin{subequations}\label{E:ls_KBF}
\begin{align}
    &\;\min_{\matr{G}} \left\|\matr{X}_2 - \matr{G}\matr{\Omega} \right\|_F^2,\,\text{where} \label{E:ls_cost} \\
    \matr{G} &= \Big[\matr{A}^d,\, \matr{B}_1^d,\, \ldots\,,\, \matr{B}_m^d\Big], \label{E:def_G}\\
    \matr{\Omega} &= \Big[\matr{X}_1^{\tran},\, \matr{\Gamma}_1^{\tran},\, \ldots\,,\, \matr{\Gamma}_m^{\tran}\Big]^{\tran}.
\end{align}
\end{subequations}

The solution to (\ref{E:ls_KBF}) is given by $\est{\matr{G}} = \matr{X}_2 \matr{\Omega}^{+}$. The pseudoinverse $\vect{\Omega}^{+}$ can be calculated from the singular value decomposition (SVD) of $\matr{\Omega} = \matr{U}\matr{\Sigma}\matr{V}^*$, where $\matr{\Sigma}$ is a diagonal matrix whose diagonal elements are the nonzero singular values in descending order, and $\matr{U}$ and $\matr{V}$ contain the corresponding left and right singular vectors. The pseudoinverse is
\begin{equation}\label{E:singular_truncation}
    \matr{\Omega}^+ = \matr{V} \matr{\Sigma}^{-1} \matr{U}^*.
\end{equation}



\subsection{Singular Value Truncation}
Since the assumed discretization of KBF is only first-order accurate w.r.t. the time step $\Delta t$ and the selected observable functions are usually not guaranteed to represent the eigenspace, using the exact pseudoinverse in the solution of (\ref{E:ls_KBF}) does not necessarily lead to optimal predictive accuracy.  Therefore the solution of (\ref{E:ls_KBF}) is modified through the technique of singular value truncation.


Assume that the number of singular values of $\matr{\Omega}$ is reduced to $r \leq q$. We get the truncated singular value matrix $\tilde{\matr{\Sigma}}\in \mathbb{R}^{r{\times} r}$ and singular vector matrices $\tilde{\matr{U}} \in \mathbb{R}^{mq{\times}r},\,\tilde{\matr{V}} \in \mathbb{R}^{N{\times}r}$. The regularized solution with singular value truncation is,
\begin{equation}
    \est{\matr{G}} = \matr{X}_2 \tilde{\matr{\Omega}}^+ = \matr{X}_2 \tilde{\matr{V}} \tilde{\matr{\Sigma}}^{-1} \tilde{\matr{U}}^*.
\end{equation}

From the composition of the parameter matrix $\matr{G}$, the singular vector matrix $\tilde{\matr{U}} = \big[\tilde{\matr{U}}_0^{\tran},\,\tilde{\matr{U}}_1^{\tran},\,\ldots\,,\,\tilde{\matr{U}}_m^{\tran}\big]^{\tran}$ is split such that $\tilde{\matr{U}}_i \in \mathbb{R}^{q{\times}r}$ to get the discretized KBF system matrices,
\begin{equation}\label{E:unreduced_KBF}
    \est{\matr{A}}^d = \matr{X}_2 \tilde{\matr{V}} \tilde{\matr{\Sigma}}^{-1} \tilde{\matr{U}}_0^*,\quad \est{\matr{B}}_i^d = \matr{X}_2 \tilde{\matr{V}} \tilde{\matr{\Sigma}}^{-1} \tilde{\matr{U}}_i^*.
\end{equation}

Assume that $\matr{P} = \matr{X}_2 \tilde{\matr{V}}$ has full column rank $r$, so that the data matrices are linearly consistent, i.e., $\mathrm{Ker}(\matr{X}_2) \subseteq \mathrm{Ker}(\tilde{\matr{\Omega}})$~\cite{tu2014}, then we can reduce the dimension of the identified discrete KBF model by taking the state transform $\bar{\vect{z}} = \matr{P}^+ \vect{z}$, which gives us the reduced KBF model,
\begin{subequations}\label{E:reduced_KBF}
\begin{align}
    \bar{\vect{z}}_{k+1} &= \bar{\matr{A}}^d \bar{\vect{z}}_k + \sum_{i=1}^m u_{i,k+1} \bar{\matr{B}}_i^d \bar{\vect{z}}_k,\,\text{where}\\
    \bar{\matr{A}}^d &= \tilde{\matr{\Sigma}}^{-1} \tilde{\matr{U}}_0^*\matr{X}_2 \tilde{\matr{V}},\\
    \bar{\matr{B}}_i^d &= \tilde{\matr{\Sigma}}^{-1} \tilde{\matr{U}}_i^*\matr{X}_2 \tilde{\matr{V}}.
\end{align}
\end{subequations}

It can be checked that the reduced system (\ref{E:reduced_KBF}) preserves the dynamics of the unreduced system (\ref{E:unreduced_KBF}) by observing that $\bar{\matr{A}}^d$ and $\est{\matr{A}}^d$ share the same nonzero eigenvalues (see~\cite{butler2015facts}) and the same is true between $\bar{\matr{B}}_i^d$ and $\est{\matr{B}}_i^d$.

In summary, the benefits for using singular value truncation include the following. (i) Reduce KBF model order in (\ref{E:reduced_KBF}); (ii) Form linear consistency between $\matr{X}_2$ and $\tilde{\matr{\Omega}}$ where the reduced $\tilde{\matr{\Omega}}$ is almost the same as $\matr{\Omega}$;  and
(iii) SVD truncation in effect adds $l_2$ regularization to the elements of $\matr{G}$~\cite{lawson1995solving}, which in discrete-time makes the solution more stable.

\subsection{Original State Reconstruction}
Since the monomial functions used as observable functions $T(\vect{x})$ contains the original state variables $\vect{x}$, the original state can be reconstructed linearly  by extracting out their entries using a constant matrix $\matr{C}^{\vect{x}}$ as in $\vect{x} = \matr{C}^{\vect{x}} \vect{z}$.


For the reduced model (\ref{E:reduced_KBF}) with the reduced observable function embedding, the hypothetical linear reconstruction takes the form $\vect{x}_{recon} = \matr{C}^{\vect{x}}\matr{P}\tilde{\vect{z}} = \matr{C}^{\vect{x}}\matr{P}\matr{P}^+\vect{z}$. However, the loss of information due to the dimension reduction of $\tilde{\vect{z}}$ may prevent the complete linear reconstruction of the original states. The information loss turns out to be the case for the Vf model in our example but not for the PQ model, and therefore the system (\ref{E:unreduced_KBF}) is used for the Vf model in order to preserve linear reconstruction.

\section{Prediction of the Whole Dynamical System}

Based on the  data-driven modules for the network, the individual models of grid-forming DERs, grid-following DERs, and power loads can be connected to form a  data-driven model for the whole microgrid system, for the prediction of its transients.
The prediction loop that connects each subsystems is provided in Algorithm~\ref{alg_prediction}, 
where the Vf and PQ controls are used for grid-forming and grid-following DER controls, respectively, and 
constant power loads are used as an example. Nonetheless, it can be easily modified to ZIP loads  or include more comprehensive load dynamics. 

Note that in order to identify an accurate KBF from data, we find that it is important to use the explicit form (\ref{E:discrete_explicit}) for the PQ controlled DER subsystem and implicit form (\ref{E:discrete_implicit}) for the Vf controlled DER subsystem. The main difference between these two DERs lies in that the Vf control has node voltage as an output while the PQ control has node voltage as an input. 
In particular, the dynamics of both types of DERs depend on their PLL state to establish the controller reference frame, which always depends on the node voltage in the same time step. Since the node voltage is an internal state in the Vf model that needs to be solved for, it follows that an implicit form need to be used. For the PQ model, the explicit form is due to the more reactive nature of the grid-following control where the node voltage is a direct input to the model.

To simplify notation, we partition all the nodes in the system into Vf node $\mathcal{P}_1$, PQ nodes with no local load $\mathcal{P}_{2a}$, PQ nodes with local load $\mathcal{P}_{2b}$, and load-only nodes $\mathcal{P}_{2c}$, after Kron reduction. Assume that the following information is given for prediction: the initial state of the Vf inverter subsystem  $\vect{x}^p_0,\, p \in \mathcal{P}_1$, initial states of the PQ inverter subsystems $\vect{x}^p_0, p \in \mathcal{P}_{2a} \cup \mathcal{P}_{2b}$, the initial mixed node current and voltage vector $\vect{u}_0$ (input to the map given by the hybrid matrix $\vect{y} = \matr{H}\vect{u}$ as in (\ref{Eq_Hmatrix})),  power loads $S_k$, and PQ inverter references $R_k$ for $k = 1,\, \ldots\,,\,PredLength$. 


The function \texttt{shiftPhase} transforms the input and output variables of each DER model between the global and local reference frames. This point is further explained in Section V-D.

\begin{algorithm}
\caption{M-KBF Enabled Prediction Algorithm}
    \label{alg_prediction}
    \SetKwInOut{KwIn}{Input}
    \SetKwInOut{KwOut}{Output}
    \SetKwFunction{shiftPhase}{shiftPhase}
    \SetKwFunction{VfModel}{VfModel}
    \SetKwFunction{PQModel}{PQModel}
    \newcommand\mycommfont[1]{\small\ttfamily\textcolor{blue}{#1}}
    \SetCommentSty{mycommfont}
    
    \KwIn{KBF models $\pi_1$ (implicit) for Vf inverter and $\pi_2$ (explicit) for PQ inverter, and hybrid network parameter matrix $\matr{H}$ for the grid}
    \KwOut{Node voltage and current in $\vect{y}_k$ and $\vect{u}_k$}
    $\vect{x}^p_0 = \shiftPhase(\vect{x}^p_0,-x_{2,0}^p)$\;
    $\vect{z}^p_0 = \matr{P}_{\pi_1}^+ T_{\pi_1}(\vect{x}^p_0),\, p \in \mathcal{P}_1$\;
    $\vect{z}^p_0 = \matr{P}_{\pi_2}^+ T_{\pi_2}(\vect{x}^p_0),\, p \in \mathcal{P}_{2a} \cup \mathcal{P}_{2b}$\;
    \For{$k \leftarrow 0$ \KwTo $PredLength$}{
        $\vect{y}_k = \matr{H}\,\vect{u}_k$\;
        $\vect{u}_{k+1} = \vect{u}_{k}$\;
        \For(\tcp*[h]{Load change}){$p \in \mathcal{P}_{2b} \cup \mathcal{P}_{2c}$}{
            $\vect{u}^p_{k+1}  \pluseq \overline{(S^p_{k+1}-S^p_k)/\vect{y}^p_k}$\;
        }
        $\vect{y}_{k+1} = \matr{H}\,\vect{u}_{k+1}$\;
         \For(\tcp*[h]{Redo load current}){$p \in \mathcal{P}_{2b} \cup \mathcal{P}_{2c}$}{
            $\vect{u}^p_{k+1} = \overline{S^p_{k+1}/\vect{y}^p_{k+1}}$\;
        }
        \For(\tcp*[h]{PQ inverter}){$p \in \mathcal{P}_{2a} \cup \mathcal{P}_{2b}$}{
            $\tilde{\vect{y}}^p_{k+1} = \shiftPhase(\vect{y}^p_{k+1},-x_{2,0}^p)$\;
            $\vect{z}^p_{k+1} = \PQModel(\vect{z}^p_k,\tilde{\vect{y}}^p_{k+1},R^p_{k+1})$\;
            $\vect{x}^p_{k+1} = \matr{C}^{\vect{x}}_{\pi_1} \matr{P}_{\pi_1} \vect{z}^p_{k+1}$\;
            \eIf{$p \in \mathcal{P}_{2b}$}{
                $\vect{u}^p_{k+1} \pluseq \shiftPhase(\matr{C} \vect{x}^p_{k+1},x_{2,0}^p)$\;
            }{
                $\vect{u}^p_{k+1} = \shiftPhase(\matr{C} \vect{x}^p_{k+1},x_{2,0}^p)$\;
            }
        }
        $\vect{y}_{k+1} = \matr{H}\,\vect{u}_{k+1}$\;
        \For(\tcp*[h]{Vf inverter}){$p \in \mathcal{P}_1$}{ 
            $\tilde{\vect{y}}^p_{k+1} = \shiftPhase(\vect{y}^p_{k+1},-x_{2,0}^p)$\;
            $\vect{z}^p_{k+1} = \VfModel(\vect{z}^p_k,\tilde{\vect{y}}^p_{k+1})$\;
            $\vect{x}^p_{k+1} = \matr{C}^{\vect{x}}_{\pi_2} \matr{P}_{\pi_2} \vect{z}^p_{k+1}$\;
            $\vect{u}^p_{k+1} = \shiftPhase(\matr{C}\vect{x}^p_{k+1},x_{2,0}^p)$\;
        }
    }
\end{algorithm}
\section{Numerical Examples}

\begin{figure}
    \centering
	\includegraphics[width=0.48\textwidth]{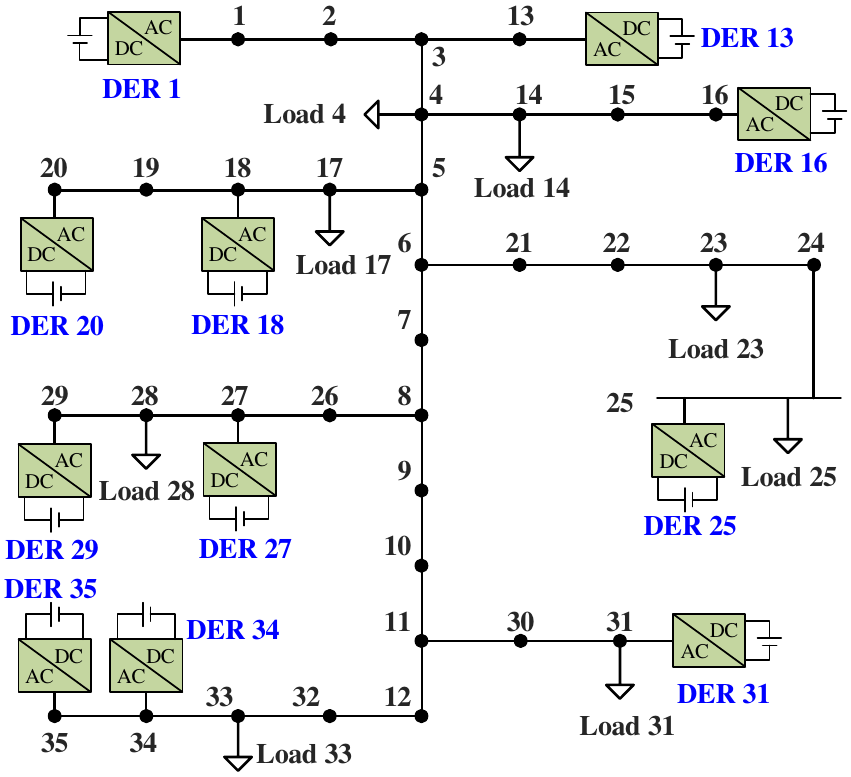}
	\caption{A typical microgrid test system}
	\label{microgrid_testsystem}
\end{figure}

A typical microgrid system shown in Fig.~\ref{microgrid_testsystem} is used to test and verify the effectiveness of the M-KBF method for modeling and predicting the nonlinear transient dynamics of the microgrid. 
The test system includes 35 buses, one Vf-controlled DER, ten PQ-controlled DERs, and eight constant power loads. For the details of the system, the reader is referred to~\cite{li2021cyber}. 
For one thing, synthetic data is prepared by modeling the test system via a set of Differential Algebraic Equations (DAEs) and simulating it through the Numerical Differenciation Formula (NDF) method with a fixed step size of $10^{-3}$ second. 
For another, simulations of the test system's transient responses to disturbances are performed to provide training data for M-KBF to identify the data-driven model and then predict the system's transient dynamics. Each transient response is induced by an approximated step change in constant power loads and/or power references of PQ-controlled DERs.
In practice, the training data set is obtained from the measurement of the controller  and the RLC filter through the advanced metering infrastructure.

Note that in the test system, the dynamical model for all PQ-controlled DERs are identical, only with  different power outputs.
It is designed to demonstrate that the identified KBF model  can be repeatedly used to efficiently identify the whole system, so as to show that the  modularity of M-KBF leads to its plug-and-play versatility. 
The M-KBF results are  analyzed from the following four aspects.

\subsection{Data-Driven Modeling for Individual DER via KBF}
\begin{figure} [!t]
    \centering
    \includegraphics[width=0.5\textwidth]{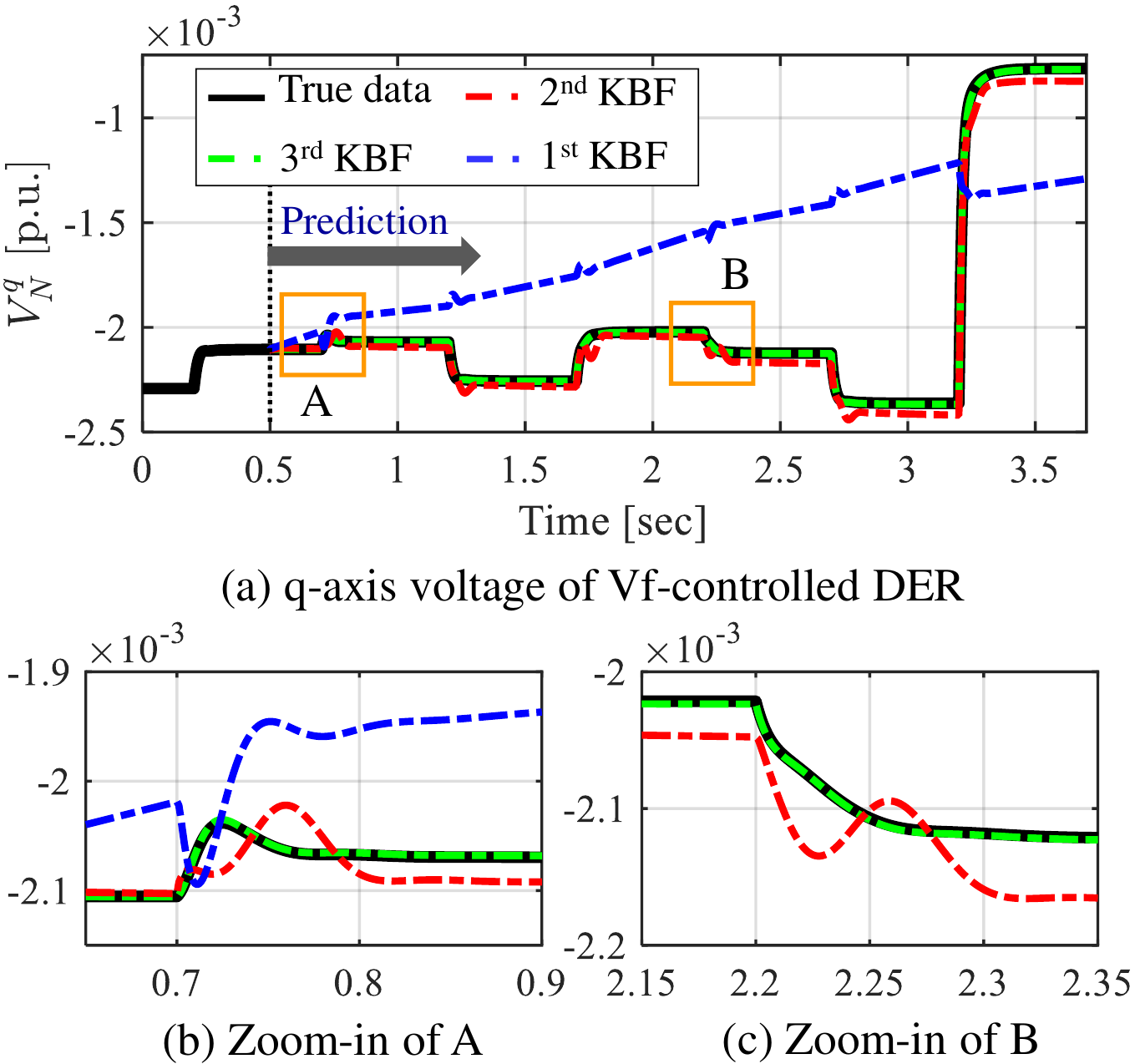}
    \caption{Prediction of Vf-controlled DER under disturbances in the $100\%$ range}
    \label{fig:monomial_order_Vf}
\end{figure}

\begin{figure}[!t]
    \centering
    \includegraphics[width=0.5\textwidth]{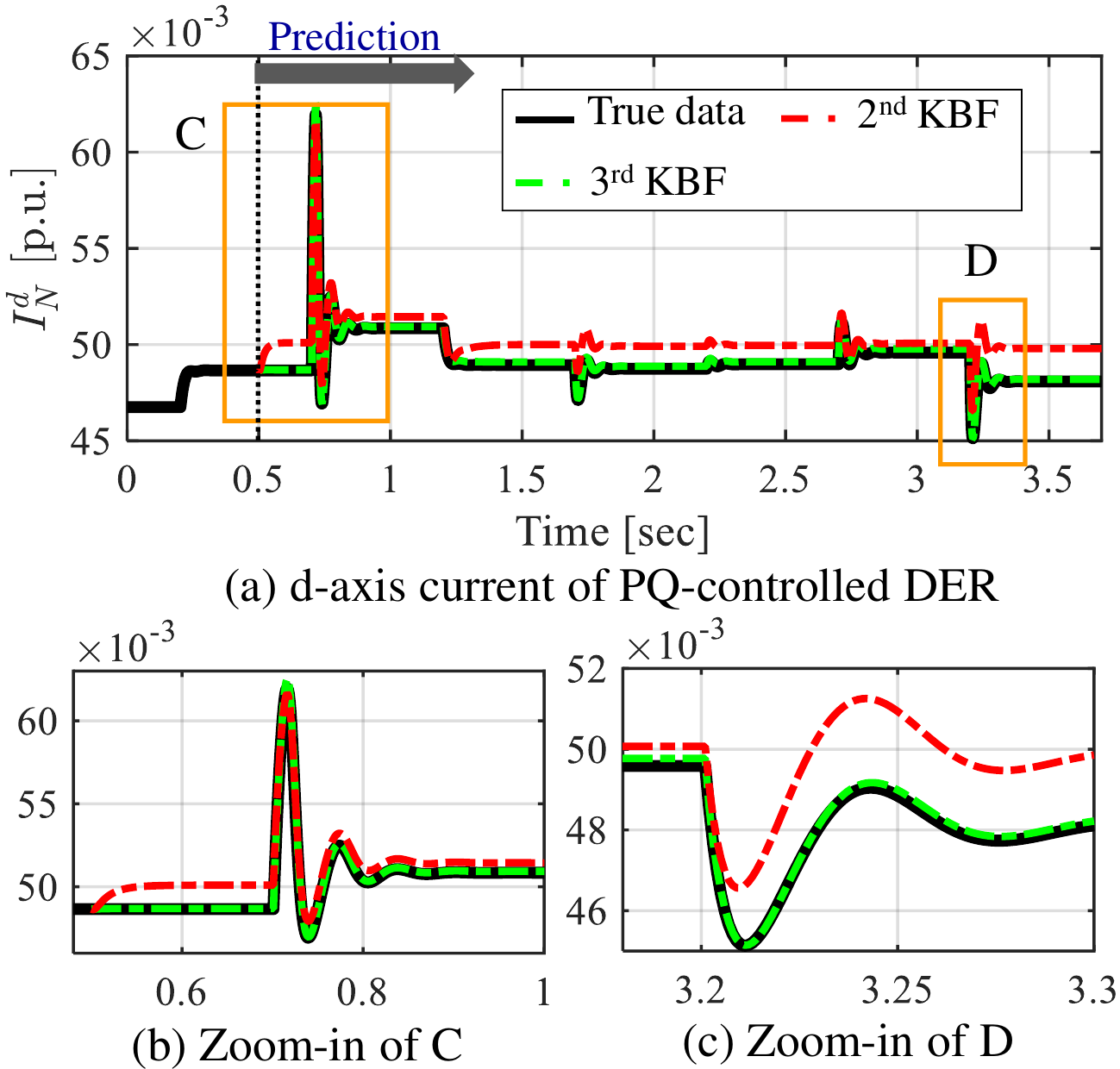}
    \caption{Prediction of PQ-controlled DER on node 31 under disturbances in the $100\%$ range}
    \label{fig:monomial_order_PQ}
\end{figure}

The synthetic training data for the Vf-controlled DER are the transient responses of the DER under  random changes of power loads and other DERs' outputs, where the changes in this paper is designed up to $\pm80\%$ of their nominal values. 
The training data for the PQ-controlled DERs are the synthetic combined transient responses of the DERs 13, 16, 18, and 20, under random changes of their power outputs that are up to $\pm20\%$ of their nominal values.
While both the $2^{\text{nd}}$-order and $3^{\text{rd}}$-order monomials can accurately predict the DER's transient responses up to $\pm80\%$ changes of power loads (or DER outputs), Fig.~\ref{fig:monomial_order_Vf} and Fig.~\ref{fig:monomial_order_PQ} show the individual model predictions in response to up to $\pm100\%$ input changes.


From the predictions, we can see that observables play an essential role in identifying the data-driven model and predicting transient dynamics, in particular when the system is under large disturbances. Detailed comparisons are:
\begin{itemize}
    \item The prediction through the KBF model obtained by the $3^{\text{rd}}$-order monomials remains accurate when the system is under $\pm100\%$  changes (note the training data is up to $\pm80\%$  changes), for both Vf- and PQ-controlled DERs.  
    \item The prediction based on the $2^{\text{nd}}$-order  monomials experiences equilibrium point drifts for Vf-controlled DER model, which causes the deviation from the true data. 
    \item The $1^{\text{st}}$-order KBF provides false prediction for the Vf-controlled DER as shown in Fig.~\ref{fig:monomial_order_Vf}; and the prediction trajectory quickly diverges for the  PQ-controlled DERs, which is too large to show in Fig.~\ref{fig:monomial_order_PQ}.
\end{itemize}

\subsection{SVD Order Truncation  and Eigenfunctions}

In order to obtain an appropriate KBF model for prediction 
and overcome the discretization error when arriving at the template model (\ref{E:discrete}),
it is necessary to apply singular value truncation to the data matrix. Here we show that an optimal truncation order can be inferred by the eigenavlue distribution. 
Fig.~\ref{fig:Vf_eigenvalue_unreduced} shows the distribution of the  approximated Koopman eigenvalues on the Vf model with monomial observables of up to $3^{\text{rd}}$ order before and after 
the optimal truncation order of $155$ is applied. We can see that: 

\begin{figure}
    \centering
    \includegraphics[width=0.5\textwidth]{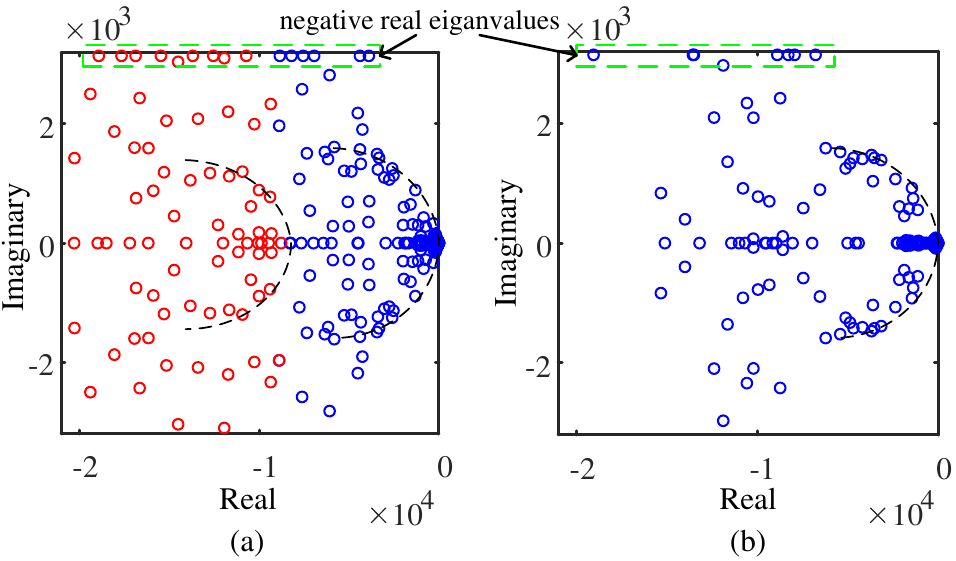}
    \caption{Identified Koopman eigenvalues before (a) and after (b) SVD truncation (Blue shows the SVD truncation order)}
    \hspace{0.1in}
    \label{fig:Vf_eigenvalue_unreduced}
\end{figure}

\begin{itemize}
    \item Before the SVD order truncation is applied to the KBF model solution, there exists an eigenvalue pattern repetition of the first $155$ eigenvalues, as illustrated by the black dashed lines in Fig.~\ref{fig:Vf_eigenvalue_unreduced}~(a). This is due to the fact that the sum of two Koopman eigenvalues is another Koopman eigenvalue with the associated eigenfunction being the product of the two existing eigenfunctions, as given in (\ref{E:eigenvalue_group}). As a consequence, the repeated eigenvalues to the left correspond to higher-order eigenfunctions that are less likely to be approximated accurately by the monomials of lower orders, which may result in an unstable KBF model for prediction.
    \item After the SVD truncation is applied, the identified Koopman eigenvalues contain no repetitive pattern and the associated eigenfunctions can be well approximated by the selected low-order monomials so that the resulting KBF model is not only stable but can accurately represent the dynamics of the DER subsystem, as has been demonstrated in Fig.~\ref{fig:monomial_order_Vf} and Fig.~\ref{fig:monomial_order_PQ}. 
\end{itemize}

We have tested Vf- and PQ-controlled DER models with both the $2^{\text{nd}}$-order and $3^{\text{rd}}$-order monomials and find that in each setup there is a narrow range of SVD truncation order that results in the KBF model having the lowest prediction error. 
The optimal truncation order is inferred by the distribution of the identified Koopman eigenvalues obtained without truncation as in Fig.~\ref{fig:Vf_eigenvalue_unreduced}~(a) by retaining only the number of eigenvalues with no repetitive pattern.

\begin{figure} 
    \centering
    \includegraphics[width=0.5\textwidth]{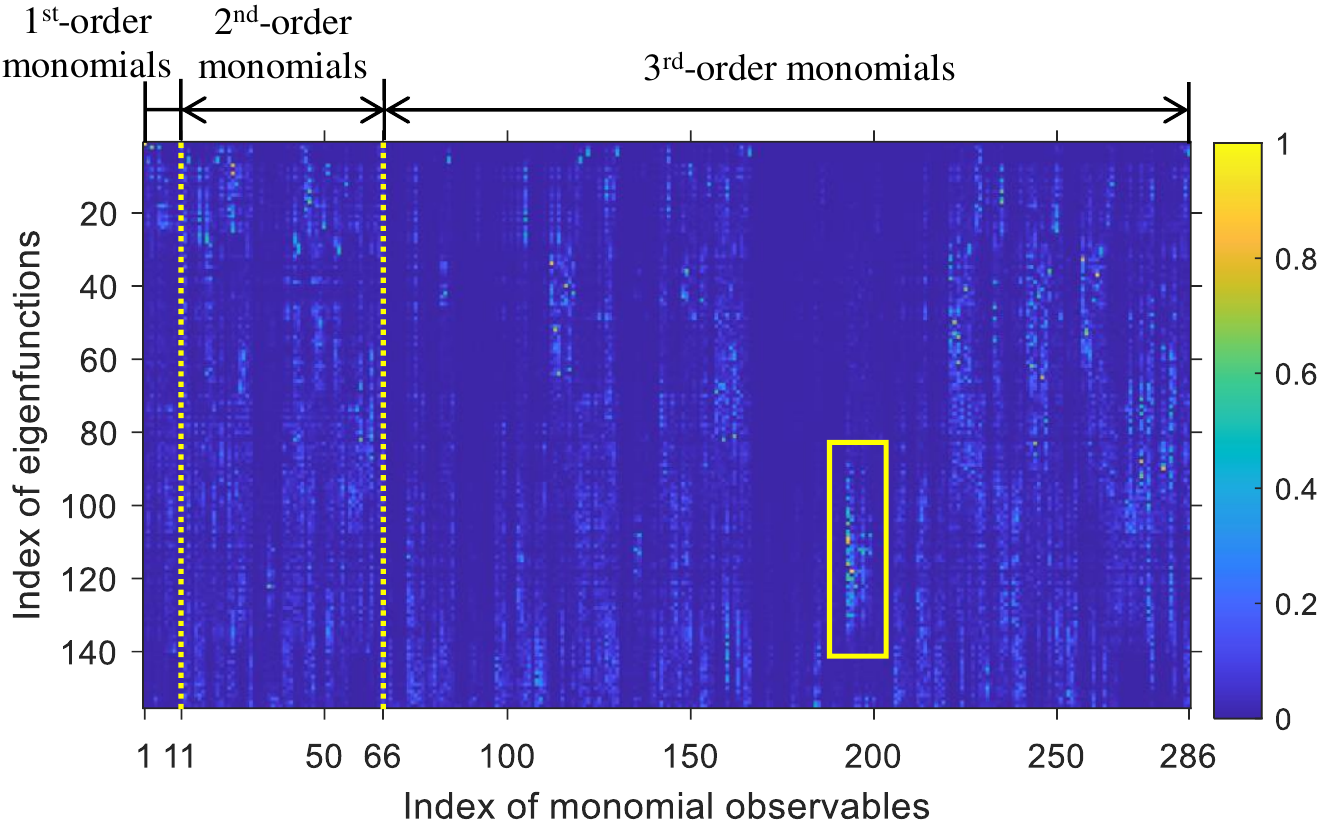}
    \caption{Eigenfunction coefficients for Vf model}
    \label{fig:Vf_eigenfunction}
\end{figure}
\begin{figure}  
    \centering
    \includegraphics[width=0.5\textwidth]{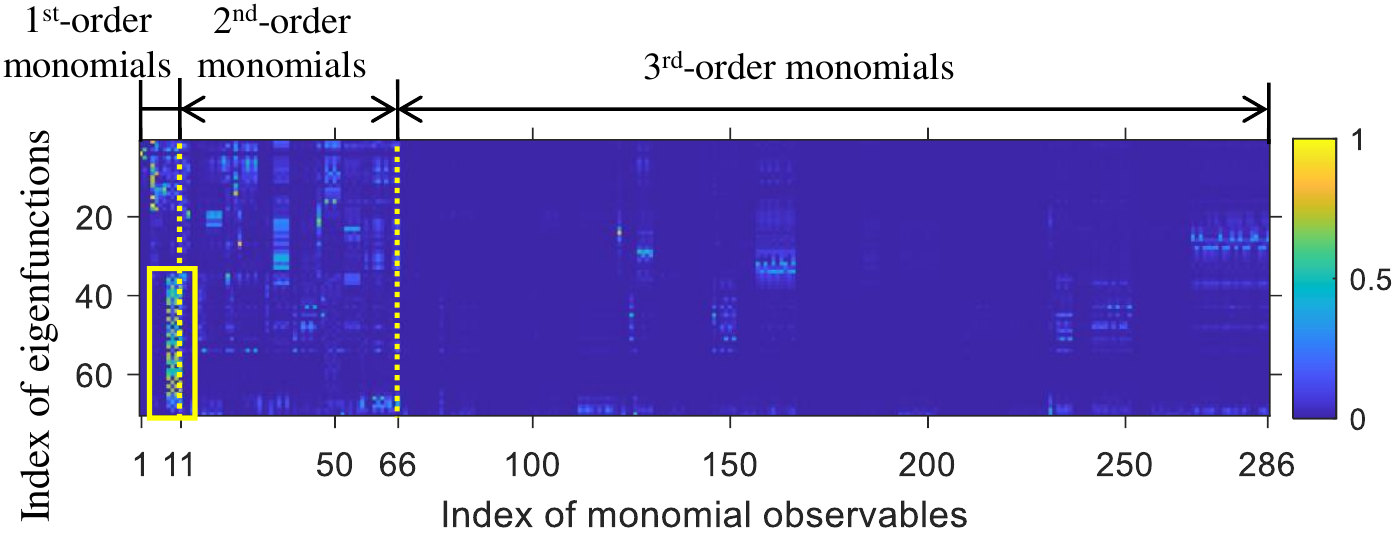}
    \caption{Eigenfunction coefficients for PQ model}
    \label{fig:PQ_eigenfunction}
\end{figure}

The coefficient matrices of the identified eigenfunction embeddings with the optimal SVD truncation orders for the Vf and PQ models are shown in Fig.~\ref{fig:Vf_eigenfunction} and Fig.~\ref{fig:PQ_eigenfunction}, where the complex coefficients are projected onto the real domain by Koopman canonical transform~\cite{surana2016koopman}.
The vertical axes in both figures correspond to the eigenfunctions in the descending order of the real parts of their eigenvalues.
The yellow rectangle in each figure shows that these eigenfunctions have large coefficients corresponding to the voltage and current variables in the DER subsystem, which agrees with the fact that these variables have fast dynamics in the solution of the DAEs. 

\subsection{Prediction based on M-KBF and Error Analysis}
\begin{figure}
    \centering
    \includegraphics[width=0.5\textwidth]{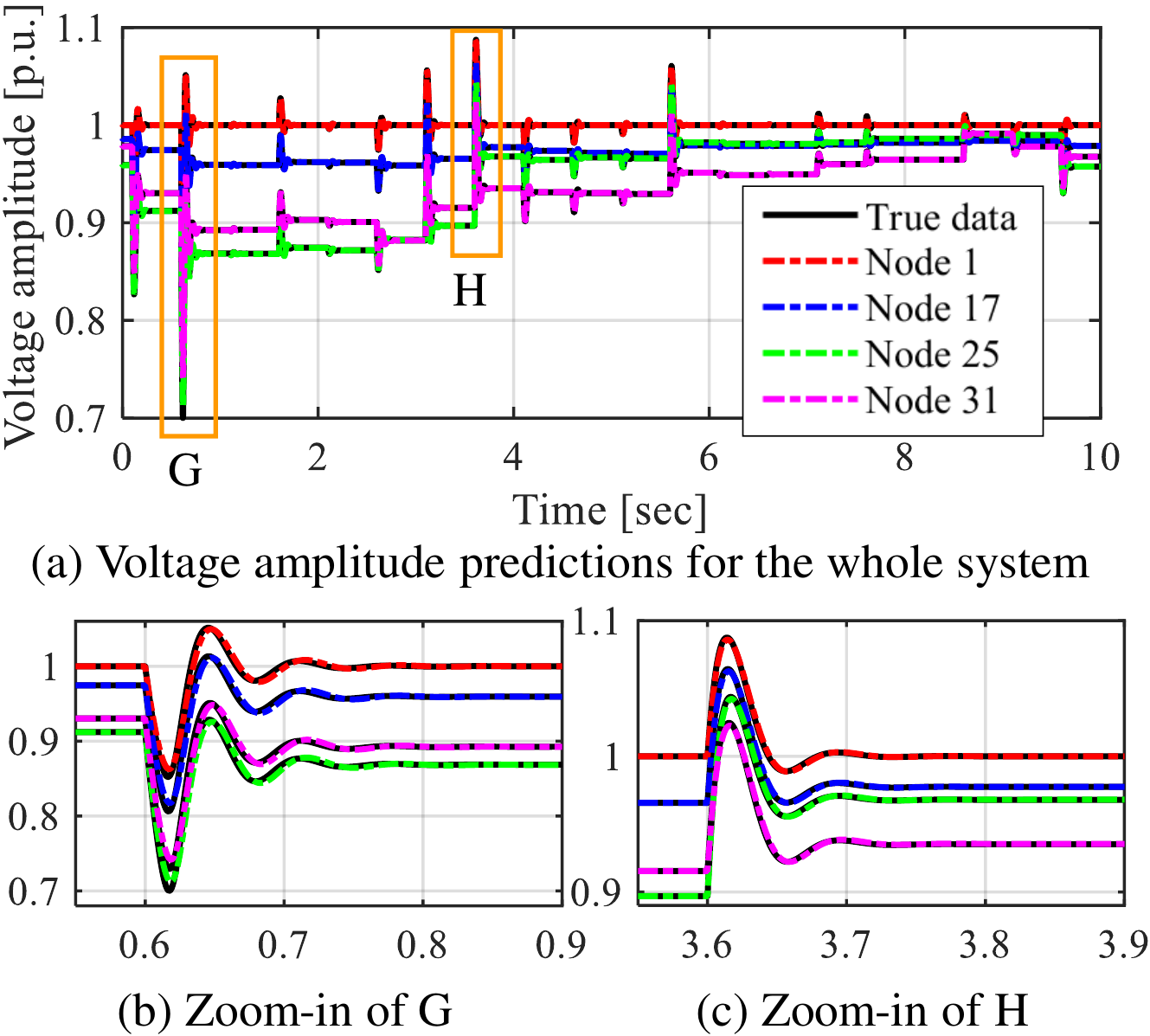}
    \caption{Voltage predictions for the whole system with $\pm100\%$ range disturbances}
    \label{fig:whole_sys_voltage}
\end{figure}

\begin{figure}
    \centering
    \includegraphics[width=0.5\textwidth]{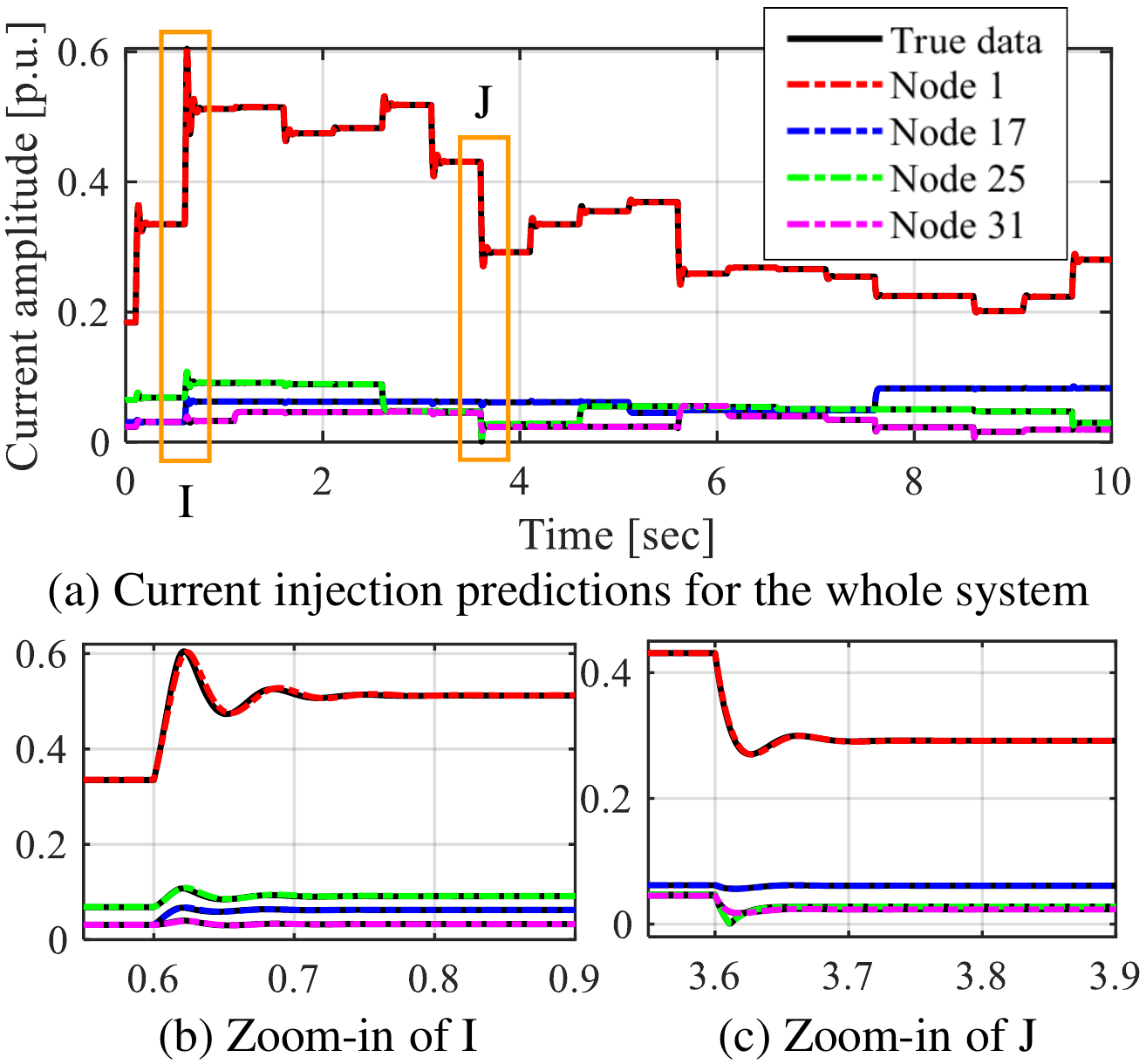}
    \caption{Current predictions for the whole system with $\pm100\%$ range disturbances}
    \label{fig:whole_sys_current}
\end{figure}

Based on Algorithm~\ref{alg_prediction}, we can integrate the identified KBF modules to perform the entire system's transient prediction. 
Fig.~\ref{fig:whole_sys_voltage} and Fig.~\ref{fig:whole_sys_current} show the voltage and current predictions of the connected M-KBF model under $\pm100\%$ range disturbances.
From the predictions, we can see that:
\begin{itemize}
    \item The Vf model trained on $\pm80\%$ range data is capable of predicting $\pm100\%$ range disturbances in the network. The PQ model trained on $\pm20\%$ range data from DERs 13, 16, 18, and 20, is capable of predicting every PQ-controlled DER in $\pm100\%$ range disturbance situations, as shown in Fig.~\ref{fig:whole_sys_voltage} and Fig.~\ref{fig:whole_sys_current}.
    \item The predictions demonstrate the M-KBF's plug-and-play function, since the identified KBF model for PQ-controlled DERs is repeatedly used in the system.
    \item The prediction error for each DER model is smaller in the connected system than that in the individual test  because each DER model's input in the connected system is a dynamic feedback from the rest of the system. 
\end{itemize}

\begin{figure}
    \centering
    \includegraphics[width=0.5\textwidth]{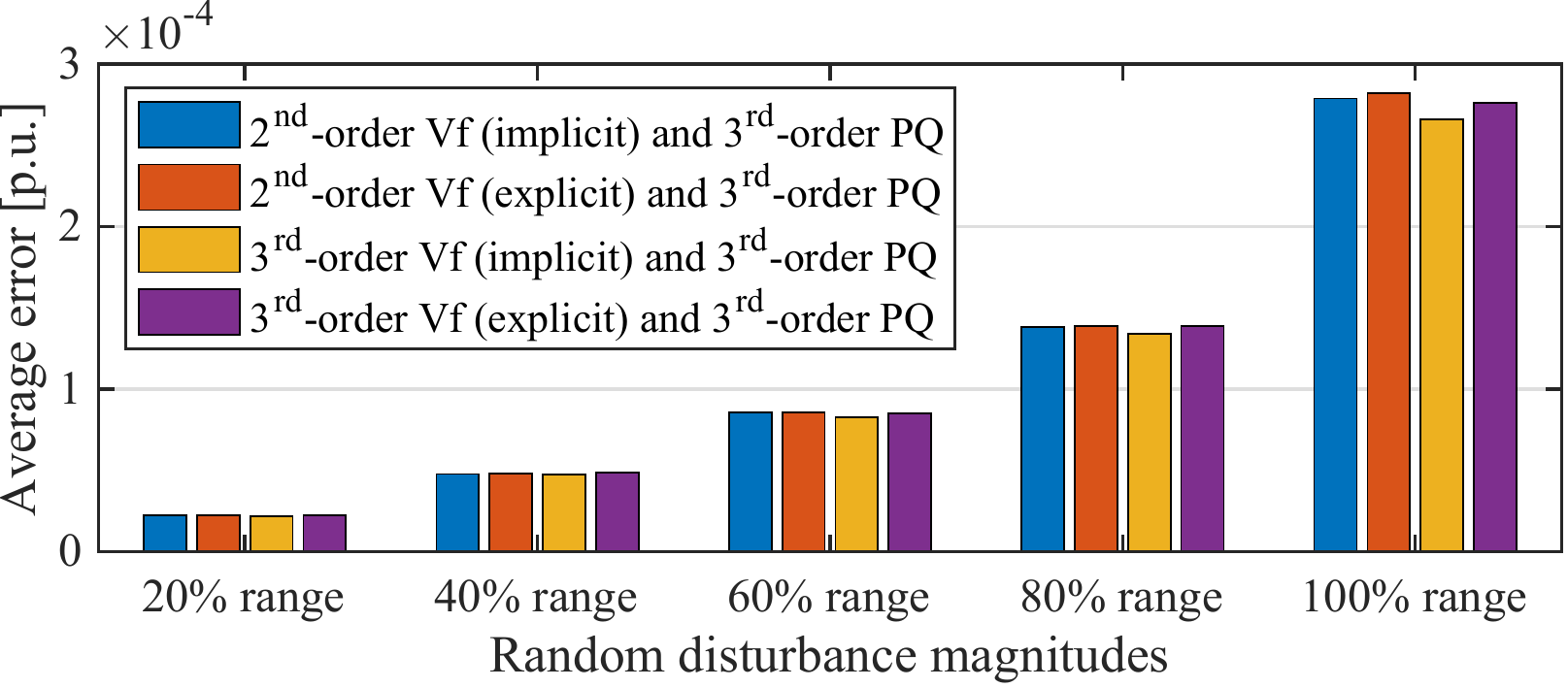}
    \caption{Average node voltage prediction errors}
    \label{fig:error_bar}
\end{figure}

The average prediction error of the M-KBF system with $2^{\text{nd}}$-order and $3^{\text{rd}}$-order monomials when the sytem is under different ranges of  disturbances  are provided in Fig.~\ref{fig:error_bar}. Overall, it shows the error increases as the disturbance increases. 
When the Vf model adopts the $3^{\text{rd}}$-order implicit prediction, we have the smallest errors compared to the $2^{\text{nd}}$-order implicit and $3^{\text{rd}}$-order explicit predictions. However, it needs a high computational effort, as summarized in Table~\ref{Table_error}, which is tested on a 2.9GHz PC.
Note that Table~\ref{Table_error} also shows that M-KBF enables faster than real-time predictions, which provides space for performing predictive control to the dynamical system.

\begin{table}[h!]
\caption{Average computational time for $20s$ prediction of the whole system with different Vf models}\label{Table_error}
\centering 
\scriptsize
\begin{tabular}{c c c c}
    \toprule[1.5pt]
    $2^{\text{nd}}$-order(explicit) &
    $2^{\text{nd}}$-order(implicit)  & 
    $3^{\text{rd}}$-order(explicit)&
    $3^{\text{rd}}$-order(implicit)   \\
    \hline
    $5.0064$s & $6.2007$s & $7.6465$s &  $20.6726$s  \\
    \bottomrule[1.5pt]
\end{tabular}
\end{table}

\subsection{Independence of KBF to the Global Reference Frame}

\begin{figure}
    \centering
    \includegraphics[width=0.5\textwidth]{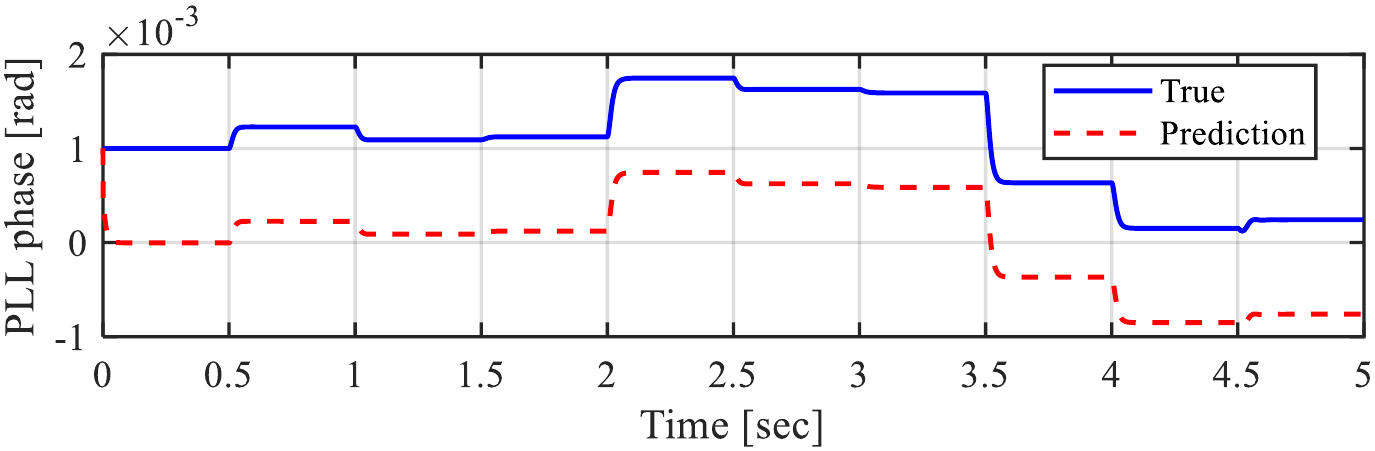}
    \caption{Counter example of dependence to DQ reference frame}
    \label{fig:reference_shift}
\end{figure}

The global DQ reference frame rotating at the nominal frequency allows one to represent the instantenous voltages and currents as phasor quantities. 
However, the initial phase of the rotating DQ frame is not unique, so that if all  phasors in the initial condition rotate by a same angle corresponding to a shift in the DQ reference frame, then the system dynamics would be the same in the new reference frame. 
This property is seen as the zero eigenvalue in the linearized system but it is a global nonlinear property, which cannot be recreated by a linear system model. 
However, the nonlinear KBF  model may still be dependent on a specific  DQ reference frame if only one frame is present in the training data. 
To illustrate this issue, Fig.~\ref{fig:reference_shift} shows the prediction of the Vf KBF model when the reference frame for the initial condition and the inputs are shifted by only $0.001$ radian. 
We can see that the predicted trajectory deviates from the initial equilibrium point and runs parallel to the simulated trajectory.

To address this issue, 
the training data is preprocessed as follows. For the Vf model, we first shift the transient responses' reference frame to zero and then randomly add an  angle within $\pm 0.01$ radian.
For the PQ model, the reference frames are only shifted to have zero  initial PLL phase angle
because the four PQ DERs are already in different reference frames.
Therefore,  the nonlinearity related to the changes between the internal PLL reference frame and the global one is represented in the KBF model in a limited range of  PLL phases.
The prediction remains accurate when the PLL phase is within a $\pm 0.04$ radian inverval for the Vf model, and $\pm 0.05$ radian interval for the PQ model. These limits could be removed by selecting proper observable functions that are invariant to different reference frames, which is the authors' future work.

\section{Conclusions}
The paper contributes a scalable data-driven method, M-KBF, to efficiently model and predict the transient dynamics of nonlinear microgrids under disturbances.
The data-driven modules for subsystems are developed through EDMD with eigenvalue-based order truncation and then integrated into a combined model for microgrid systems.
Test results have demonstrated the effectiveness of M-KBF in providing fast and precise transient predictions. 
For future work, M-KBF will be further developed to handle data with noise, to identify the system by using output-only measurements like bus voltage and current, and to perform predictive control based on the predictions obtained from M-KBF.






\appendix

\begin{figure}
    \centering
    \includegraphics[width=0.5\textwidth]{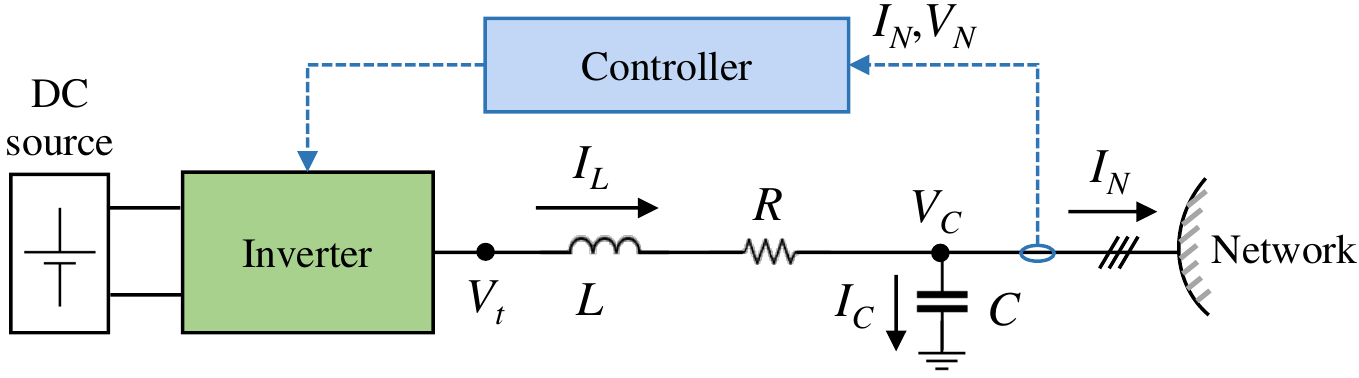}
    \caption{The connection circuit of DER}
    \label{fig_connection_diagram}
\end{figure}

\begin{figure}
    \centering
    \includegraphics[width=0.5\textwidth]{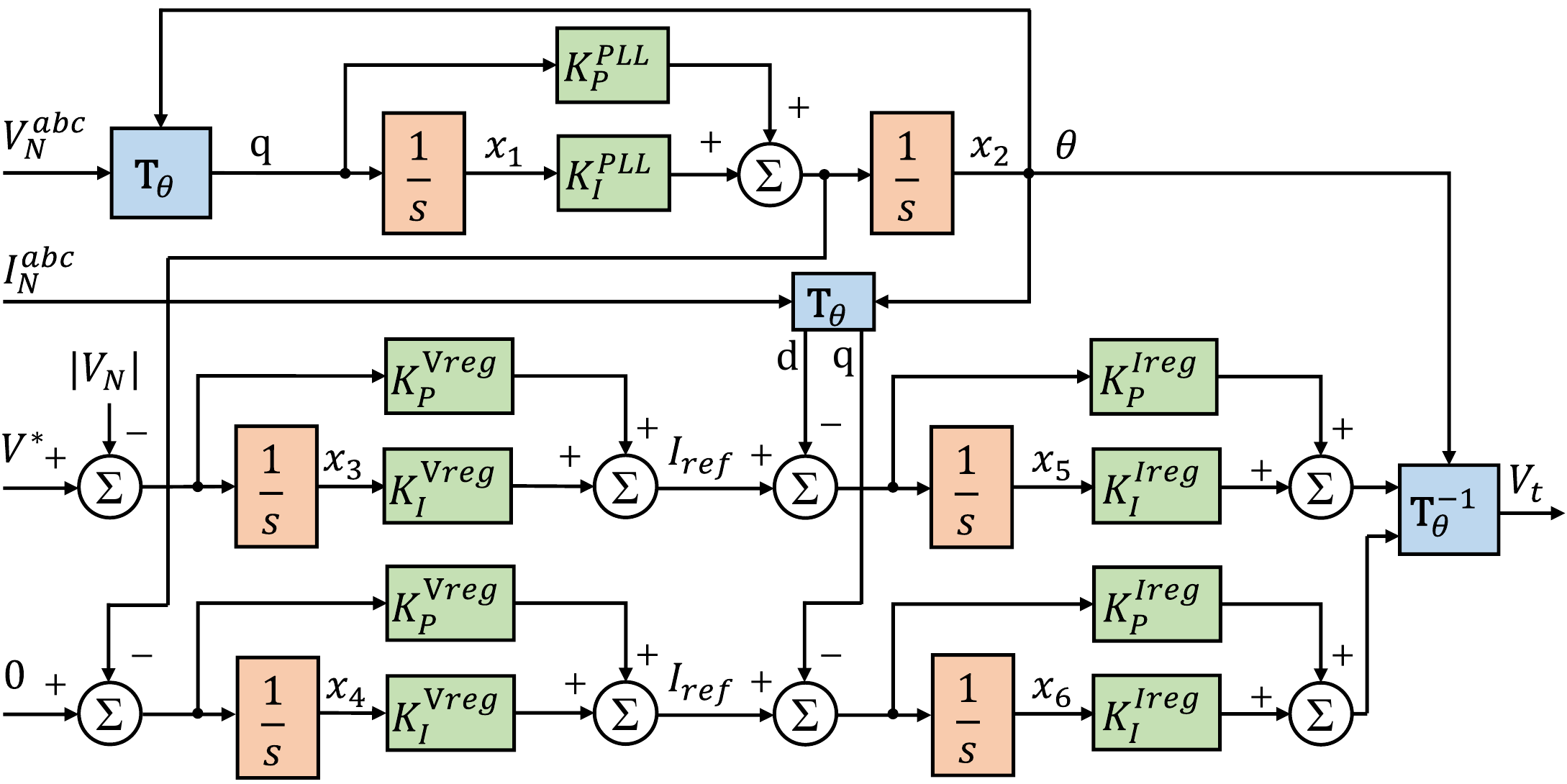}
    \caption{Vf control}
    \label{fig_Vf_block_diagram}
\end{figure}

The DER connection circuit and the Vf-controller (given as  an example) are shown in Fig.~\ref{fig_connection_diagram} and Fig.~\ref{fig_Vf_block_diagram}, respectively.
Ten differential equations can then be developed, which are not given here due to the page limit.
These equations form a control affine system, where the state vector is $\vect{x} = \big[x_1,\,x_2,\,x_3,\,x_4,\,x_5,\,x_6,\,V_C^D,\,V_C^Q,\,I_L^D,\,I_L^Q\big]^{\tran}$, the input vector is the node current $\vect{u} = \big[I_N^D,\,I_N^Q\big]^{\tran}$, and the output to the  network is the node voltage that is equal to the state variables $\vect{y} = \big[V_C^D,\,V_C^Q\big]^{\tran}$. 

The input functions  of the control affine system, $\vect{g}_1(\vect{x})$ and $\vect{g}_2(\vect{x})$, associated with each input  are  given by,
\begin{align}
    \vect{g}_1 &= \Big[\,0,\,0,\,0,\,0,\,-\cos{x_2},\,\sin{x_2},\,\frac{1}{C},\,0,\,-\frac{K_p^{Ireg}}{L},\,0\,\Big]^{\tran}, \\
    \vect{g}_2 &= \Big[\,0,\,0,\,0,\,0,\,-\sin{x_2},\,-\cos{x_2},\,0,\,\frac{1}{C},\,0,\,-\frac{K_p^{Ireg}}{L}\,\Big]^{\tran}.
\end{align}



\bibliographystyle{IEEEtran}
\bibliography{References}

\begin{thebibliography}{10}
\providecommand{\url}[1]{#1}
\csname url@samestyle\endcsname
\providecommand{\newblock}{\relax}
\providecommand{\bibinfo}[2]{#2}
\providecommand{\BIBentrySTDinterwordspacing}{\spaceskip=0pt\relax}
\providecommand{\BIBentryALTinterwordstretchfactor}{4}
\providecommand{\BIBentryALTinterwordspacing}{\spaceskip=\fontdimen2\font plus
\BIBentryALTinterwordstretchfactor\fontdimen3\font minus
  \fontdimen4\font\relax}
\providecommand{\BIBforeignlanguage}[2]{{%
\expandafter\ifx\csname l@#1\endcsname\relax
\typeout{** WARNING: IEEEtran.bst: No hyphenation pattern has been}%
\typeout{** loaded for the language `#1'. Using the pattern for}%
\typeout{** the default language instead.}%
\else
\language=\csname l@#1\endcsname
\fi
#2}}
\providecommand{\BIBdecl}{\relax}
\BIBdecl

\bibitem{sobbouhi2021transient}
A.~R. Sobbouhi and A.~Vahedi, ``Transient stability prediction of power system;
  a review on methods, classification and considerations,'' \emph{Electric
  Power Systems Research}, vol. 190, p. 106853, 2021.

\bibitem{li2021determination}
X.~Li, C.~Mishra, S.~Chen, Y.~Wang, and J.~De~La~Ree, ``Determination of
  parameters of time-delayed embedding algorithm using koopman operator-based
  model predictive frequency control,'' \emph{CSEE Journal of Power and Energy
  Systems}, vol.~7, no.~6, pp. 1140--1151, 2021.

\bibitem{adibi2019reinforcement}
M.~Adibi and J.~van~der Woude, ``A reinforcement learning approach for
  frequency control of inverted-based microgrids,'' \emph{IFAC-PapersOnLine},
  vol.~52, no.~4, pp. 111--116, 2019.

\bibitem{madani2021data}
S.~S. Madani and A.~Karimi, ``Data-driven lpv controller design for islanded
  microgrids,'' \emph{IFAC-PapersOnLine}, vol.~54, no.~7, pp. 433--438, 2021.

\bibitem{a299a76f179d4ebba7c9872e1aced639}
J.~Sanchez-Gasca, D.~Trudnowski, E.~Barocio, J.~Hauer, J.~Pierre, C.~Canizares,
  H.~Huang, T.~Rauhala, J.~Chow, I.~Kamwa, J.~Turunen, M.~Crow, G.~Ledwich,
  L.~Vanfretti, L.~Dosiek, R.~Martin, V.~Vittal, H.~Ghasemi, E.~Martinez,
  D.~Vowles, M.~Gibbard, A.~Messina, R.~Wies, L.~Haarla, B.~Pal, and N.~Zhou,
  \emph{\BIBforeignlanguage{English}{Identification of Electromechanical Modes
  in Power Systems}}.\hskip 1em plus 0.5em minus 0.4em\relax United States:
  IEEE, 2012.

\bibitem{119247}
J.~Hauer, ``Application of prony analysis to the determination of modal content
  and equivalent models for measured power system response,'' \emph{IEEE
  Transactions on Power Systems}, vol.~6, no.~3, pp. 1062--1068, 1991.

\bibitem{5275444}
D.~J. Trudnowski and J.~W. Pierre, ``Overview of algorithms for estimating
  swing modes from measured responses,'' in \emph{2009 IEEE Power Energy
  Society General Meeting}, 2009, pp. 1--8.

\bibitem{260898}
I.~Kamwa, R.~Grondin, E.~Dickinson, and S.~Fortin, ``A minimal realization
  approach to reduced-order modelling and modal analysis for power system
  response signals,'' \emph{IEEE Transactions on Power Systems}, vol.~8, no.~3,
  pp. 1020--1029, 1993.

\bibitem{780912}
J.~Sanchez-Gasca and J.~Chow, ``Performance comparison of three identification
  methods for the analysis of electromechanical oscillations,'' \emph{IEEE
  Transactions on Power Systems}, vol.~14, no.~3, pp. 995--1002, 1999.

\bibitem{6024892}
L.~L. Grant and M.~L. Crow, ``Comparison of matrix pencil and prony methods for
  power system modal analysis of noisy signals,'' in \emph{2011 North American
  Power Symposium}, 2011, pp. 1--7.

\bibitem{4410548}
G.~Liu, J.~Quintero, and V.~M. Venkatasubramanian, ``Oscillation monitoring
  system based on wide area synchrophasors in power systems,'' in \emph{2007
  iREP Symposium - Bulk Power System Dynamics and Control - VII. Revitalizing
  Operational Reliability}, 2007, pp. 1--13.

\bibitem{9609618}
A.~F. El~Hamalawy, M.~Ammar, H.~F. Sindi, M.~F. Shaaban, and H.~H. Zeineldin,
  ``A subspace identification technique for real time stability assessment of
  droop based microgrids,'' \emph{IEEE Transactions on Power Systems}, pp.
  1--1, 2021.

\bibitem{6982236}
E.~Barocio, B.~C. Pal, N.~F. Thornhill, and A.~R. Messina, ``A dynamic mode
  decomposition framework for global power system oscillation analysis,''
  \emph{IEEE Transactions on Power Systems}, vol.~30, no.~6, pp. 2902--2912,
  2015.

\bibitem{stoica2005spectral}
P.~Stoica, R.~L. Moses \emph{et~al.}, \emph{Spectral analysis of
  signals}.\hskip 1em plus 0.5em minus 0.4em\relax Pearson Prentice Hall Upper
  Saddle River, NJ, 2005.

\bibitem{anderson2005bootstrap}
M.~G. Anderson, N.~Zhou, J.~W. Pierre, and R.~W. Wies, ``Bootstrap-based
  confidence interval estimates for electromechanical modes from multiple
  output analysis of measured ambient data,'' \emph{IEEE Transactions on Power
  Systems}, vol.~20, no.~2, pp. 943--950, 2005.

\bibitem{liu2008oscillation}
G.~Liu and V.~Venkatasubramanian, ``Oscillation monitoring from ambient pmu
  measurements by frequency domain decomposition,'' in \emph{2008 IEEE
  International Symposium on Circuits and Systems}.\hskip 1em plus 0.5em minus
  0.4em\relax IEEE, 2008, pp. 2821--2824.

\bibitem{8598677}
H.~N. Villegas~Pico, B.~Mather, and G.-S. Seo, ``Model identification of
  inverter nonlinear control dynamics,'' in \emph{2018 IEEE Electronic Power
  Grid (eGrid)}, 2018, pp. 1--6.

\bibitem{1546078}
R.~Fard, M.~Karrari, and O.~Malik, ``Synchronous generator model identification
  for control application using volterra series,'' \emph{IEEE Transactions on
  Energy Conversion}, vol.~20, no.~4, pp. 852--858, 2005.

\bibitem{brunton2016discovering}
S.~L. Brunton, J.~L. Proctor, and J.~N. Kutz, ``Discovering governing equations
  from data by sparse identification of nonlinear dynamical systems,''
  \emph{Proceedings of the national academy of sciences}, vol. 113, no.~15, pp.
  3932--3937, 2016.

\bibitem{mezic2005spectral}
I.~Mezi{\'c}, ``Spectral properties of dynamical systems, model reduction and
  decompositions,'' \emph{Nonlinear Dynamics}, vol.~41, no.~1, pp. 309--325,
  2005.

\bibitem{mezic2004comparison}
I.~Mezi{\'c} and A.~Banaszuk, ``Comparison of systems with complex behavior,''
  \emph{Physica D: Nonlinear Phenomena}, vol. 197, no. 1-2, pp. 101--133, 2004.

\bibitem{mezic2021koopman}
I.~Mezic, ``Koopman operator, geometry, and learning of dynamical systems,''
  \emph{Notices of the AMS}, vol.~68, no.~7, pp. 1087--1105, 2021.

\bibitem{BEVANDA2021197}
\BIBentryALTinterwordspacing
P.~Bevanda, S.~Sosnowski, and S.~Hirche, ``Koopman operator dynamical models:
  Learning, analysis and control,'' \emph{Annual Reviews in Control}, vol.~52,
  pp. 197--212, 2021. [Online]. Available:
  \url{https://www.sciencedirect.com/science/article/pii/S1367578821000729}
\BIBentrySTDinterwordspacing

\bibitem{budivsic2020koopman}
M.~Budi{\v{s}}ic, R.~Mohr, and I.~Mezic, ``The koopman operator in systems and
  control: Concepts, methodologies, and applications,'' 2020.

\bibitem{susuki2011nonlinear}
Y.~Susuki and I.~Mezic, ``Nonlinear koopman modes and coherency identification
  of coupled swing dynamics,'' \emph{IEEE Transactions on Power Systems},
  vol.~26, no.~4, pp. 1894--1904, 2011.

\bibitem{korda2018power}
M.~Korda, Y.~Susuki, and I.~Mezi{\'c}, ``Power grid transient stabilization
  using koopman model predictive control,'' \emph{IFAC-PapersOnLine}, vol.~51,
  no.~28, pp. 297--302, 2018.

\bibitem{sharma2021data}
P.~Sharma, V.~Ajjarapu, and U.~Vaidya, ``Data-driven identification of
  nonlinear power system dynamics using output-only measurements,'' \emph{IEEE
  Transactions on Power Systems}, 2021.

\bibitem{williams2015data}
M.~O. Williams, I.~G. Kevrekidis, and C.~W. Rowley, ``A data--driven
  approximation of the koopman operator: Extending dynamic mode
  decomposition,'' \emph{Journal of Nonlinear Science}, vol.~25, no.~6, pp.
  1307--1346, 2015.

\bibitem{folkestad2020extended}
C.~Folkestad, D.~Pastor, I.~Mezic, R.~Mohr, M.~Fonoberova, and J.~Burdick,
  ``Extended dynamic mode decomposition with learned koopman eigenfunctions for
  prediction and control,'' in \emph{2020 american control conference
  (acc)}.\hskip 1em plus 0.5em minus 0.4em\relax IEEE, 2020, pp. 3906--3913.

\bibitem{korda2020optimal}
M.~Korda and I.~Mezi{\'c}, ``Optimal construction of koopman eigenfunctions for
  prediction and control,'' \emph{IEEE Transactions on Automatic Control},
  vol.~65, no.~12, pp. 5114--5129, 2020.

\bibitem{goswami2021bilinearization}
D.~Goswami and D.~A. Paley, ``Bilinearization, reachability, and optimal
  control of control-affine nonlinear systems: A koopman spectral approach,''
  \emph{IEEE Transactions on Automatic Control}, 2021.

\bibitem{peitz2020data}
S.~Peitz, S.~E. Otto, and C.~W. Rowley, ``Data-driven model predictive control
  using interpolated koopman generators,'' \emph{SIAM Journal on Applied
  Dynamical Systems}, vol.~19, no.~3, pp. 2162--2193, 2020.

\bibitem{kettner2021harmonic}
A.~M. Kettner, L.~Reyes-Chamorro, J.~K.~M. Becker, Z.~Zou, M.~Liserre, and
  M.~Paolone, ``Harmonic power-flow study of polyphase grids with
  converter-interfaced distributed energy resources—part i: Modeling
  framework and algorithm,'' \emph{IEEE Transactions on Smart Grid}, vol.~13,
  no.~1, pp. 458--469, 2021.

\bibitem{kettner2017properties}
A.~M. Kettner and M.~Paolone, ``On the properties of the power systems nodal
  admittance matrix,'' \emph{IEEE Transactions on Power Systems}, vol.~33,
  no.~1, pp. 1130--1131, 2017.

\bibitem{dorfler2012kron}
F.~Dorfler and F.~Bullo, ``Kron reduction of graphs with applications to
  electrical networks,'' \emph{IEEE Transactions on Circuits and Systems I:
  Regular Papers}, vol.~60, no.~1, pp. 150--163, 2012.

\bibitem{7384725}
A.~Mauroy and I.~Mezić, ``Global stability analysis using the eigenfunctions
  of the koopman operator,'' \emph{IEEE Transactions on Automatic Control},
  vol.~61, no.~11, pp. 3356--3369, 2016.

\bibitem{bruce2019koopman}
A.~L. Bruce, V.~M. Zeidan, and D.~S. Bernstein, ``What is the koopman operator?
  a simplified treatment for discrete-time systems,'' in \emph{2019 American
  Control Conference (ACC)}.\hskip 1em plus 0.5em minus 0.4em\relax IEEE, 2019,
  pp. 1912--1917.

\bibitem{proctor2016dynamic}
J.~L. Proctor, S.~L. Brunton, and J.~N. Kutz, ``Dynamic mode decomposition with
  control,'' \emph{SIAM Journal on Applied Dynamical Systems}, vol.~15, no.~1,
  pp. 142--161, 2016.

\bibitem{tu2014}
J.~H.~Tu, C.~W.~Rowley, D.~M.~Luchtenburg, S.~L.~Brunton, and J.~Nathan~Kutz,
  ``On dynamic mode decomposition: Theory and applications,'' \emph{Journal of
  Computational Dynamics}, vol.~1, no.~2, p. 391–421, 2014.

\bibitem{butler2015facts}
D.~Butler, ``Facts about eigenvalues,'' \emph{Textbook, University of
  California, San diego, US}, 2015.

\bibitem{lawson1995solving}
C.~L. Lawson and R.~J. Hanson, \emph{Solving least squares problems}.\hskip 1em
  plus 0.5em minus 0.4em\relax SIAM, 1995.

\bibitem{li2021cyber}
Y.~Li, \emph{Cyber-Physical Microgrids}.\hskip 1em plus 0.5em minus 0.4em\relax
  Springer, 2021.

\bibitem{surana2016koopman}
A.~Surana, ``Koopman operator based observer synthesis for control-affine
  nonlinear systems,'' in \emph{2016 IEEE 55th Conference on Decision and
  Control (CDC)}.\hskip 1em plus 0.5em minus 0.4em\relax IEEE, 2016, pp.
  6492--6499.

\end{thebibliography}
\end{document}